\documentclass[journal]{IEEEtran}
\newcommand{\reals}{{\mathrm{I\kern-.2em R}}}
\newcommand{\complex}{{\mathrm{C\kern-.6em C}}}
\newcommand{\field}{{\mathrm{I\kern-.2em F}}}
\newcommand{\expectation}{{\mathrm{I\kern-.2em E}}}

\newcommand{\argmax}{\mathop{\rm arg\ max}}

\newcommand{\dd }{{\rm d}}

\newcommand{\tr}{{\rm tr}}

\newcommand{\expect}{{\rm E}}

\newcommand{\calK}{{\cal K}}

\newcommand{\calM}{{\cal M}}

\newcommand{\lp}{l^\prime}

\newcommand{\np}{n^\prime}

\newcommand{\pp}{p^\prime}

\newcommand{\qp}{q^\prime}

\newcommand{\thetap}{\theta^\prime}
\newcommand{\phip}{\phi^\prime}

\newcommand{\zetap}{\zeta^\prime}

\newcommand{\vzero}{{\bf 0}}

\newcommand{\vc}{{\bf c}}

\newcommand{\vg}{{\bf g}}

\newcommand{\vk }{{\bf k}}

\newcommand{\vx }{{\bf x}}

\newcommand{\vv }{{\bf v}}

\newcommand{\vw }{{\bf w}}
\newcommand{\vy }{{\bf y}}

\newcommand{\vmu }{{\mbox{\boldmath$\mu$}}}

\usepackage{lineno}
\usepackage{times,amsfonts,amsmath,mathtools,epsfig,fixltx2e,amsthm,amssymb}
\usepackage{mathtools,amssymb,nccmath}
\usepackage{graphics,graphicx,epstopdf}
\usepackage{url}
\usepackage{cite}
\newcommand{\withpro}{$\mbox{PhI}^{{\rm Pro}+}$}
\newcommand{\withoutpro}{$\mbox{PhI}^{{\rm Pro}-}$}
\newcommand{\C}{C}
\newcommand{\CrhoREAL}{\C} 
\newcommand{\CcREAL}{\C_{p_1,\zeta_1;p_2,\zeta_2}} 
\newcommand{\CcprimeREAL}{\C_{p_1^\prime,\zeta_1^\prime;p_2^\prime,\zeta_2^\prime}} 
\newcommand{\Crhorho}{\C^{\rho,\rho}}
\newcommand{\Crhorhoconj}{\C^{\rho,\rho^\ast}}
\newcommand{\Ccc}{\C_{p_1,\zeta_1;p_2,\zeta_2}^{c,c}}
\newcommand{\Cccconj}{\C_{p_1,\zeta_1;p_2,\zeta_2}^{c,c^\ast}}
\newcommand{\CRecRec}{\C_{p_1,\zeta_1;p_2,\zeta_2}^{\Re c,\Re c}}
\newcommand{\CRecImc}{\C_{p_1,\zeta_1;p_2,\zeta_2}^{\Re c,\Im c}}
\newcommand{\CImcImc}{\C_{p_1,\zeta_1;p_2,\zeta_2}^{\Im c,\Im c}}
\newcommand{\nonnegative}{\reals^+\cup\{0\}}
\newcommand{\hetero}{{\tt Hetero}}
\newcommand{\heterohomo}{{\tt Hetero} {\em (zero covar)}} 
\newcommand{\heteroasymmetry}{{\tt Hetero} {\em (Asymmetry)}}
\newcommand{\heterosymparticles}{{\tt Hetero} {\em (SymPart)}} 
\newcommand{\heterosymstatistics}{{\tt Hetero} {\em (SymStat)}} 

\newcommand{\Nirrep}{N_{\rm rep}}
\newcommand{\Ngroup}{N_g}

\begin{document}
\title{Reconstruction of stochastic 3-D signals with symmetric statistics from 2-D projection images motivated by cryo-electron microscopy}
\author{Nan~Xu,~\IEEEmembership{Member, IEEE}
        and~Peter~C.~Doerschuk,~\IEEEmembership{Senior Member, IEEE}
\thanks{N. Xu was with the School
of Electrical and Computer Engineering, Cornell University, Ithaca, NY, 14853 USA.  She is now with the Wallace H. Coulter Department of Biomedical Engineering, Georgia Institute of Technology and Emory University, Atlanta, Georgia 30332, USA e-mail: im.nan.xu@gmail.com
}
\thanks{P. C. Doerschuk is with the School
	of Electrical and Computer Engineering and the Meinig School
	of Biomedical Engineering, Cornell University, Ithaca, NY, 14853 USA e-mail: pd83@cornell.edu}
}
%
\markboth{IEEE Transactions on Image Processing}{Xu and Doerschuk: Symmetric Statistical Characterization of Heterogeneous Virus Particles}
\maketitle
\begin{abstract}
Cryo-electron microscopy provides 2-D projection images of the 3-D electron
scattering intensity of many instances of the particle under study (e.g., a
virus).
Both symmetry (rotational point groups) and heterogeneity are important
aspects of biological particles and both aspects can be combined by
describing the electron scattering intensity of the particle as a
stochastic process with a symmetric probability law and therefore symmetric
moments.
A maximum likelihood estimator implemented by an expectation-maximization
algorithm is described which estimates the unknown statistics of the
electron scattering intensity stochastic process from images of instances
of the particle.
The algorithm is demonstrated on the bacteriophage HK97 and the virus
N$\omega$V.
The results are contrasted with existing algorithms which assume that each
instance of the particle has the symmetry rather than the less restrictive
assumption that the probability law has the symmetry.
\end{abstract}
%
\IEEEpeerreviewmaketitle
\section{Introduction}
\label{sec:introduction}
\IEEEPARstart{S}{tructural Biology} is the study of the 3-D geometry
of biological particles such as viruses, ribosomes, enzymes, {\em
  etc}.
The 3-D geometry, typically called the ``structure,'' can be as detailed as
the 3-D location of each atom in the particle or, if only lower resolution
is achievable, the electron scattering intensity as a function of
position in 3-D.
Understanding the 3-D geometry of synthetic nanoparticles is also becoming
important in materials
science~\cite{MaGongAubertTurkerKaoDoerschukWiesnerNature2018}.
\par
Single-particle cryo-electron microscopy
(cryo-EM)~\cite{JensenMethodsEnzymologyABC2010} is an experimental method
for structure determination which is of great and increasing
importance~\cite{SubramaniamKuhlbrandtHendersonRecentAdvancesInCryoEMIUCrJ2016}
and which has been recognized with the 2017 Nobel Prize in
Chemistry~\cite{nobelprizechemistry2017}.
In this method,
many instances of the particle are flash frozen to the temperature of
liquid nitrogen.
Images with pixel dimensions of about 1 {\AA} are recorded.
Each image shows many instances of the particle.
No instance of the particle is imaged more than once in order to minimize
damage to the instance of the particle from the electron beam of the
microscope.
The many images (typically $10^4$--$10^6$), each a projection image with an
unknown projection direction, are computationally combined to
yield one 3-D reconstruction of the electron scattering intensity of the
particle.
In favorable situations, nearly atomic resolution (e.g., 2--3 {\AA})
spatial resolution can be achieved in the 3-D reconstruction of the
electron scattering intensity.
This paper describes an integrated approach to two aspects of this
reconstruction problem: heterogeneity among the particles and symmetry of
the particles.
\par
Alternative techniques involving electron microscopy include electron
crystallography (based on 2-D crystals) and electron cryotomography (based
on recording images of a single specimen at a variety of tilt positions of
the microscope's stage, generally limited to approximately $\pm 70^\circ$
of tilt)~\cite{Cheng.ReviewCryoEM.Science.2018}.
The idea of symmetry constraints on the statistics of the electron
scattering intensity, which is central to this paper, could be applied to
these alternative methods.
However, the estimation problem relating the data and the electron
scattering intensity would be different for each of these alternative
methods.
\subsection{Heterogeneity}
\label{sec:introduction:heterogeneity}
Instances of a particle can differ, i.e., the particles are heterogeneous.
Two mechanisms by which heterogeneity can occur are compositional
differences and flexibility.
An eukaryotic 80S ribosome can be an example of a particle with
compositional differences.
Such a ribosome
(i) has a molecular mass of about $3.2\times 10^6$ Dalton,
(ii) has dimensions of 250--300~{\AA}, and
(iii) is composed of about 80 protein molecules plus about 4 RNA molecules.
The ribosome's function is to assemble on the mRNA that will be translated
into the corresponding protein molecule and catalyze the translation.
Such a ribosome has been reported to have compositional
variability~\cite{SlavovSemrauAiroldiBudnikVanOudenaardenRibosomeHeterogeneityCellReports2015}
and such variability is one source of heterogeneity.
\par
Flexibility is a second source of heterogeneity, as is
indicated by the fact that the protein structures obtained from X-ray
crystallography experiments do not always show the entire protein amino
acid sequence (which is separately known from molecular biology methods)
but rather lacks certain disordered portions of the molecule.
When heterogeneity is due to particle motions, at least large-space
long-time characteristics of the motion are preserved in the frozen
specimen that is imaged because the freezing to the temperature of liquid
nitrogen is fast ($10^6$ ${}^\circ$C/sec with vitrification occurring in
$10^{-4}$
sec~\cite{DubochetAdrianChangHomoLepaultMcDowallSchultzEarlyCryoEMQuarterlyRevBiophys1988}).
\par
As is described in the second paragraph of Section~\ref{sec:introduction},
the image shows many instances of the particle.
Subimages showing individual instances are extracted and used for further
processing.
This extraction process has errors which complicate characterizing the
heterogeneity of the instances.
\par
When all instances of the particle are identical, standard software (e.g.,
Refs.~\cite{TangPengBaldwinMannJiangReesLudtkeJSB2007EMAN2,ScheresRELIONimplementationJMB2012,Shaikh...Frank.spider.NatureProtocols2008})
can typically compute a structure.
When each instance of the particle comes from one of a few number of
classes (typically less than ten) and all instances in a class are
identical (described as discrete heterogeneity), standard software (e.g.,
Refs.~\cite{TangPengBaldwinMannJiangReesLudtkeJSB2007EMAN2,ScheresRELIONimplementationJMB2012,Shaikh...Frank.spider.NatureProtocols2008})
can often compute a structure for each class.
\par
The case when the instances in a class differ (described as continuous
heterogeneity) has been described as an important issue in
multiple review articles over a range of years, e.g.,
\cite[p.~221]{TaylorGlaeserJSB2008} to
\cite[p.~55]{BaiMcMullanScheres2015} to
the report of the 2017 Nobel Prize in
Chemistry~\cite{nobelprizechemistry2017}.
Describing the heterogeneity of the instances of a particle by a mixture
density with a finite number of classes~\cite{RednerWalker1984} is a
combination of discrete and continuous heterogeneity.
If only a single class is used then the continuous heterogeneity might
describe large scale variability among the instances of the particle.
Alternatively, if a finite set of classes is used then the changes among
the instances within a class might be smaller.
Multiple statements of this problem exist and multiple algorithms exist
which address the solution of the different problem
statements~\cite{Jonic.conformationalheterogeneity.CurrOpinStructBiol.2017}.
The problem statement addressed in this paper is (1)~to characterize
continuous heterogeneity within a class directly from the image data without
additional information, (2)~to provide a statistical characterization that
is relevant for smaller-scale fluctuations that would occur within a class,
and (3)~to apply symmetry constraints to the statistics rather than to the
individual instances.
\par
Alternative problem statements focus on one or more issues including the
following issues:
\begin{itemize}
\item
Use additional information:
For instance,
Refs.~\cite{JinSorzanoRosaTrevinBilbaoCastroNunezRamirezLlorcaTamaJonic.normalmodes.Structure.2014,SorzanoRosaTrevinTamaJonic.normalmodes.JSB.2014}
start with a reference structure from which normal modes can be computed
and then compare projections of the reference structure deformed by the
normal modes with the images.
\item
Large-scale motions:
For instance,
Refs.~\cite{DashtiETCFrankOurmazdRibosomeBrownianMachinePNAS2014,FrankOurmazd.Methods.2016}
is a manifold-based method for interpolating among the many structures that
can be computed from certain datasets such as the ribosome dataset of
Ref.~\cite{DashtiETCFrankOurmazdRibosomeBrownianMachinePNAS2014}.
\item
Compute properties of the
continuously heterogeneous particle ensemble:
For instance, Ref.~\cite{TagareKucukelbirSigworthWangRao.JSB.2015}
computes, directly from image data, the first few (e.g., 5) principal
values and components of the 3-D electron scattering intensity of the
particle.
\end{itemize}
\par
Many problem statements and algorithm solutions, some of which are
described above, have been recently
reviewed~\cite{Jonic.conformationalheterogeneity.CurrOpinStructBiol.2017}.
Additional recent
reviews~\cite{Ludtke.MethodsEnzymology.2016,Scheres.MethodsEnzymology.2016}
by the primary authors of two of the widely used software systems describe
fewer problem statements and algorithm solutions.
For instance, Ref.~\cite{Ludtke.MethodsEnzymology.2016} describes only
methods for dealing with discrete heterogeneity and
Ref.~\cite{Scheres.MethodsEnzymology.2016} describes only methods for
dealing with discrete heterogeneity with some extensions to some forms of
continuous heterogeneity when subregions of the structure move as rigid
bodies~\cite[Sections~4.4--4.6]{Scheres.MethodsEnzymology.2016}.
Therefore, new methods for characterizing continuous heterogeneity are needed.
\par
Data resampling~\cite{PenczekKimmelSpahn2011,SpahnPenczek2009,ZhangKimmelSpahnPenczek2008,PenczekYangFrankSpahnBootstrapMethodJSB2006,SimonettiMarziMyasnikovFabbrettiYusupovGualerziKlaholzNature2008} is one class of existing approaches:
multiple datasets are created by resampling the original dataset, a
structure is computed for each dataset, and then statistics are computed by
averaging over the multiple structures.
When the particle has symmetry, the symmetry is generally imposed on each
of the structures.
This may be the reason that the resulting spatial variance functions (i.e.,
the variance as a function of position) are known to have anomalous peaks
on and near symmetry axes of the
particle~\cite[p.~173]{Ludtke.MethodsEnzymology.2016}.
Since the symmetry axes can be the location of important biology, e.g., the
Flock House Virus example of Section~\ref{sec:introduction:symmetry},
anomalous results near symmetry axes is not a satisfactory situation.
However, computing asymmetry reconstructions and allowing the symmetry to
appear in the averaging operation requires substantially more data and
computation and may not be practical.
\par
In order to describe a second class of approaches, based on
moments~\cite{LiaoHashemFrankCovarianceCryoEMStructure2015,LiaoFrankIEEEISBI2010,KatsevichKatsevichSingerCovarianceCryoEMSIAMImagingSciences2015,AndenSinger.SIAMImagingScience.2018},
it is helpful to describe the mathematical description of the image that is
used.
Different methods describe the instance of the particle using different
mathematics, for instance, a 3-D array of voxel
values~\cite{ScheresRELIONimplementationJMB2012}, a truncated 3-D Fourier
series~\cite{DoerschukJohnsonIT2000}, or a weighted sum of basis functions
where the basis functions are chosen in order to achieve sparse matrix
operations~\cite{KatsevichKatsevichSingerCovarianceCryoEMSIAMImagingSciences2015}.
The voxel values, Fourier series coefficients, or weights for the $i$th
particle (or, equivalently, the $i$th image since each particle is imaged
only once) are packed into a vector which is denoted by $c_i$.
The $i$th image, denoted by $y_i$, is a linear transformation of $c_i$ plus
an additive noise, denoted by $w_i$.
The transformation, denoted by $L$, includes physical processes such as the
3-D to 2-D projection and the Contrast Transfer Function (CTF) of the
microscope.
The transformation $L$ contains parameters, e.g., the direction of the 3-D
to 2-D projection, which vary from particle to particle and these
parameters are denoted by $\theta_i$.
Therefore, $y_i=L(\theta_i)c_i + w_i$.
The collections
$\{\theta_i\}|_{i=1}^{N_v}$,
$\{c_i\}|_{i=1}^{N_v}$,
$\{w_i\}|_{i=1}^{N_v}$ are independent and the
$\theta_i$,
$c_i$ (mean $\bar c$, covariance $\mathbf{V}$), and
$w_i$ (mean $0$, covariance $\mathbf{Q}$)
are all i.i.d.
Therefore, conditional on the value of $\theta_i$, the mean and covariance of
$y$ is
$\bar y=L(\theta_i)\bar c$
and
$\Sigma=L(\theta_i)\mathbf{V}L^T(\theta_i) + \mathbf{Q}$,
respectively.
Idealizations in the mathematical model include the fact that multiple
noise sources are present and the CTF of the microscope effects some of the
noise sources, the noise is not purely additive, and the noise is not
independent of the signal.
\par
Having described the mathematical model, we can describe the second class
of approaches which are based on 
moments~\cite{LiaoHashemFrankCovarianceCryoEMStructure2015,LiaoFrankIEEEISBI2010,KatsevichKatsevichSingerCovarianceCryoEMSIAMImagingSciences2015,AndenSinger.SIAMImagingScience.2018}.
The basic idea of the moment estimators is to estimate $\Sigma$ by a sample
covariance and then estimate $\mathbf{V}$ from the covariance equation which is
linear in $\mathbf{V}$.
All methods must assume that the values of $\theta_i$, which are estimated from
the data, are correct.
Additionally, if the instances of the particle come from multiple classes
of particle, the estimate of the class must also be correct.
Resolution is limited in some methods, e.g., $16\times 16\times 16$
voxels~\cite{LiaoHashemFrankCovarianceCryoEMStructure2015}.
In some
methods~\cite{AndenKatsevichSingerISBI2015,KatsevichKatsevichSingerCovarianceCryoEMSIAMImagingSciences2015},
the solution of the linear system is done in a basis where the system is
sparse thereby enabling the solution of large problems, e.g., $10^3\times 10^3$.
Estimation of $\theta_i$ would probably be done in the context of assuming
that the covariance $\mathbf{V}$ is zero.
Ref.~\cite[Supplemental Figure~1]{YunyeGongVeeslerDoerschukJohnsonJSB2016}
provides some numerical results on the extent to which estimates of
$\theta_i$ change when the covariance $\mathbf{V}$ is simultaneously
estimated.
In the particular case of
Ref.~\cite{YunyeGongVeeslerDoerschukJohnsonJSB2016}, about 10\% of the
particles changed orientation, including 5\% with large orientation
changes.
\par
Using a model similar to the model of the previous paragraph, the mean and
covariance of the random vectors $c_i$ can be estimated for each class by a
third approach which uses a
maximum likelihood estimator computed by an expectation-maximization
algorithm in which $\theta_i$ are the nuisance
parameters~\cite{YiliZhengQiuWangDoerschukJOSA2012,QiuWangMatsuiDomitrovicYiliZhengDoerschukJohnsonJSB2012,TangKearneyQiuWangDoerschukBakerJohnsonJMolRecog2014,DomitrovicMovahedBothnerMatsuiQiuWangDoerschukJohnsonJMB2013,YunyeGongVeeslerDoerschukJohnsonJSB2016,DoerschukGongXuDomitrovicJohnsonCurOpinVirology2016,XuVeeslerDoerschukJohnsonHK97JSB2017}.
The present paper is an extension of these ideas, which are described in
more detail in Section~\ref{sec:MLreconstruction}.
\par
The biological analysis is typically based on the variance function.
In more detail, let the electron scattering stochastic process be denoted
by $\rho(\vx)\in\mathbb{R}^1$ where $\vx\in\mathbb{R}^3$ and expectation be denoted by
$\expect$.
Then the mean function is $\bar\rho(\vx)=\expect[\rho(\vx)]$, the
covariance function is
$C(\vx_1,\vx_2)=\expect[(\rho(\vx_1)-\bar\rho(\vx_1))(\rho(\vx_2)-\bar\rho(\vx_2))]$,
the variance function is $v(\vx)=C(\vx,\vx)$, and the standard
deviation function is $s(\vx)=\sqrt{v(\vx)}$ which is usefully primarily
because it has the same units as $\rho(\vx)$.
Focusing on the variance function, which is the common choice, clearly
ignores a great deal of critically important information
since $C(\vx_1,\vx_2)$ contains information on how the electron
density behaves at spatially separated locations, but that information is
missing from $v(\vx)$.
To the best of our knowledge, this is the first publication in which a
complete 6-D covariance ($C(\vx_1,\vx_2)$) analysis is performed for
particles with symmetry.
In the numerical examples of Section~\ref{sec:results} the covariance
matrix $\mathbf{V}$ that determines $C(\vx_1,\vx_2)$ is assumed to be
diagonal in order to decrease the amount of computation but all the
formulas in the paper apply for arbitrary $\mathbf{V}$.
\subsection{Symmetry}
\label{sec:introduction:symmetry}
An important characteristic of many systems is symmetry, i.e.,
the electron scattering intensity is invariant under the operations of a
symmetry group.
On 05 November 2017, the Protein Data Bank~\cite{proteindatabank} contained
130,005 structures in seven symmetry classes:
Asymmetric (60.198\%),
Cyclic $C_n$ (31.020\% with $n$ ranging from 2 to 39~\cite{CasanasQuerolAudiGuerraPousTanakaTsukiharaVerdaguerFitaVaultParticleC39ActaCrystD2013,TanakaKatoYamashitaSumizawaZhouYaoIwasakiYoshimuraTsukiharaVaultParticleC39Science2009}
),
Dihedral $D_n$ (7.517\% with $n$ ranging from 2 to 48~\cite{AndersonKickhoeferSieversRomeEisenbergVaultShell9AngstromD48PLoSBiol2007}),
Icosahedral $I$ (0.491\%),
Tetrahedral $T$ (0.280\%),
Octahedral $O$ (0.248\%),
and Helical (0.245\%).
We focus on the rotational point groups, which are $I$, $O$, $T$, $C_n$,
and $D_n$, and which have 60, 24, 12, $n$, and $2n$ symmetry operations,
respectively.
Helical is a line group.
%
\par
Many viruses have a protein shell that surrounds
the genome of the virus and the shell has icosahedral
symmetry~\cite{RossmannJohnsonIcosahedralVirusAnnualReviewBiochem1989}.
The most compelling evidence for the existence of such symmetries is
atomic-resolution x-ray crystallography.
In particular,
the particle has sufficiently small compositional heterogeneity and
flexibility heterogeneity to form a crystal;
the non-physiological particle-particle contacts in the crystal further
reduce heterogeneity;
high-resolution (e.g., 2.0--3.0~{\AA}) x-ray diffraction data is recorded;
the data is analyzed under the assumption of icosahedral symmetry;
and the resulting atomic positions agree with chemical (e.g., bond
lengths), biochemical (e.g., left-handed amino acids), and molecular
biology (e.g., sequence of amino acids) knowledge about the particle which
is the evidence that the icosahedral symmetry assumption used in the
analysis of the data is correct.
But even particles with atomic-resolution x-ray crystallographic structures
are not rigid.
For instance, Flock House Virus has such a structure, the pentamer of
$\gamma$ peptides near the 5-fold symmetry axis of the icosahedrally
symmetric particle appear on the interior surface of the capsid in the
atomic-resolution
structure~\cite{FisherJohnson.FHV.crystalstructure.Nature.1993,Cheng...BakerJohnson.FHV.crystallographyandcryoEM.Structure.1994},
but biochemical and cell biological evidence indicates that the pentamer is
sometimes on the surface of the particle and plays a role in the entry of
the particle into a new host
cell~\cite{Odegard...Johnson.FHV.gamma.CurrentTopicsMicrobiolImmun.2010,Walukiewicz...Johnson.FHV.gamma.JVirology.2008}.
\par
In this paper, we assume that the correct symmetry is known.
If there is uncertainty in the symmetry, then multiple calculations can be
performed, each with a different symmetry, and the results (including the
value of the likelihood function and whether the answer makes sense to the
biologist) can be employed to determine the correct symmetry.
\subsection{Combining symmetry and heterogeneity and outline of the paper}
\label{sec:combiningsymmetryandheterogeneity}
\label{sec:outline}
In this paper we combine ideas of symmetry and continuous
heterogeneity by applying the symmetry to the statistics of the particle.
This implies (Section~\ref{sec:symmetrysymmetricstatistics})
that the moments of the electron scattering intensity satisfy constraints.
Our numerical results appear to indicate that this approach avoids
introducing anomalous peaks in the variance function ($v(\vx)$) on and
near symmetry axes of the particle which occur with standard resampling
methods of estimating the variance
function~\cite[p.~173]{Ludtke.MethodsEnzymology.2016}, as was described in
Section~\ref{sec:introduction:heterogeneity}.
\par
Details of the symmetry groups influence the organization of the paper.
In particular, if all the irreducible representations~\cite[Theorem~3.3
  p.~69]{Miller1972} (see also paragraph~3 of
Section~\ref{sec:BasisFunction}) (irreps) of the group
are real, then there are suitable real-valued basis functions so that an
orthonormal expansion of the real-valued electron scattering intensity can be
determined with real-valued coefficients.
Real irreps are possible for the icosahedral group, which is the symmetry
group of our example virus reconstruction problems, and so the main body of
the paper focuses on that simpler case while the case where real irreps are
not possible, which includes important cases such as cyclic symmetry, is
described in Appendix~\ref{sec:constraintsonthemomentsoftheweights:complex}.
\par
The choice of the coordinate system in $\mathbb{R}^3$ that is used to describe
the particle also influences the organization of the paper.
Particles exhibiting $I$, $O$, or $T$ symmetries are likely to be most
parsimoniously described in spherical coordinates but particles exhibiting
$C_n$ or $D_n$ may be better described in cylindrical coordinates.
Since our example has icosahedral ($I$) symmetry, the main body of the
paper focuses on spherical coordinates while the case of cylindrical
coordinates is described in Appendix~\ref{sec:nonsphericalcoordinatesystems}.
\par
The reminder of the paper is organized as follows.
The maximum likelihood estimator
approach~\cite{YiliZhengQiuWangDoerschukJOSA2012} that is generalized in
this paper with a new signal model is described in
Section~\ref{sec:MLreconstruction}.
The new signal model is described in parts:
symmetry (Section~\ref{sec:symmetrysymmetricstatistics}),
basis functions respecting the symmetry (Section~\ref{sec:BasisFunction})
and
constraints on the linear combinations of basis functions needed to achieve
symmetric statistics (Section~\ref{sec:constraintsonthemomentsoftheweights}
and Appendices~\ref{sec:mean:derivation}, \ref{sec:covariance:derivation}, and
\ref{sec:finitedimensional}).
The changes needed in the maximum likelihood estimator in order
to use the new signal model are described in Section~\ref{sec:MLE}.
Numerical results on simulated and experimental data are described in
Section~\ref{sec:results} and conclusions in Section~\ref{sec:conc}.
\subsection{Notation}
\label{sec:notation}
\begin{enumerate}
\item
``Probability density function'' (``probability mass function'') is
abbreviated by ``pdf'' (``pmf'').
\item
The multivariable Gaussian pdf with mean $\mu$ and covariance $\Sigma$
evaluated at $y$ is denoted by $N(\mu,\Sigma)(y)$.
\item
$\expect$ denotes expectation.
\item
i.i.d.\ stands for ``independent and identically distributed.''
\item
$\vx\in\mathbb{R}^3$ and $x=\|\vx\|_2$.
\item
The quantity $\vx/x$ is a unit vector and, since $\vx\in\mathbb{R}^3$, is
shorthand for the two angles of spherical coordinates.
\item
Superscript ${}^\ast$ is complex conjugation.
\item
$I_n$ is the $n\times n$ identity matrix and $0_{m,n}$ is the $m\times n$
zero matrix.
\item
$\int \dd\Omega$ is integration over the surface of the sphere in 3-D,
which, in spherical coordinates, is $\int_{\theta=0}^\pi
\int_{\phi=0}^{2\pi} \sin(\theta)\dd\theta\dd\phi$.
If $\vx\in\mathbb{R}^3$ is described in spherical coordinates as
$\vx=(x,\theta,\phi)$ and $f$ is a function of $\vx$ then $\int
f(\vx)\dd\Omega$ is a function of $x$, i.e., the integral is over the
surface of a sphere at radius $x$.
\item
The Kronecker delta function $\delta_{m,n}$ is defined by $\delta_{m,n}=1$ if
$m=n$, and $\delta_{m,n}=0$ otherwise, where $m$ and $n$ are integers or vectors of integers.
\end{enumerate}
\section{Maximum likelihood reconstruction}
\label{sec:MLreconstruction}
Because this paper extends Ref.~\cite{YiliZhengQiuWangDoerschukJOSA2012},
in this section we summarize Ref.~\cite{YiliZhengQiuWangDoerschukJOSA2012}
for the special case where all the particles that are imaged come from the
same class and where the pdf on the projection orientations is known.
We describe the problem in the context of cryo-EM, but any application
having a linear imaging equation containing nusiance parameters would share
the equations.
\par
The electron scattering intensity $\rho(\vx)$ is described as a linear
superposition of known basis functions $F_\omega(\vx)$ with weights
$c_\omega$:
\begin{equation}
\rho(\vx)
=
\sum_{\omega\in\Omega}
c_\omega
F_\omega(\vx)
\label{eq:rho:general}
.
\end{equation}
The goal of the reconstruction is to estimate the joint pdf on the
$c_\omega$.
\par
The imaging process is linear from $\rho(\vx)$ to image and therefore also
from $c_\omega$ to image, with unknown nuisance parameters in the linear
transformation, and has additive noise.
Let $\vy_i$ be the $i$th image.
Let $\vc_i$ be the vector of $c_\omega$ for the $i$th image.
Let $z_i$ be the nuisance parameters for the $i$th image and
$\mathbf{L}(z_i)$ be the linear
transformation~\cite{Erickson1973,LepaultPitt1984,ToyoshimaUnwin1988,Scherzer1949}.
Let $\vw_i$ be the noise.
Then~\cite[Eq.~3]{YiliZhengQiuWangDoerschukJOSA2012},
\begin{equation}
\vy_i=\mathbf{L}(z_i)\vc_i+\vw_i
.
\label{eq:2Dmodel}
\end{equation}
The random variables $z_i$, $\vc_i$, $\vw_i$ ($i\in\{1,\dots,N_v\}$) are
independent and
$z_i$ is uniform on rotations,
$\vc_i\sim N(\bar{\vc},\mathbf{V})$, and
$\vw_i\sim N(\vzero,\mathbf{Q})$.
\par
The maximum likelihood estimator (MLE) of $\bar{\vc}$ and $\mathbf{V}$ is
computed by a generalized expectation-maximization (E-M)
algorithm
using the $z_i$ as nuisance variables.
The update equation for the new value of $\bar{\vc}$
is~\cite[Eq.~33]{YiliZhengQiuWangDoerschukJOSA2012}
$\mathbf{F} \bar{\vc}^{\rm new}=\vg$ where
\begin{eqnarray}
\mathbf{F}
&=&
\sum_{i=1}^{N_v}
\int_{z_i}
\!\!
\mathbf{L}^T(z_i)
\mathbf{\Sigma}_i^{-1}(z_i,\mathbf{V}_0)
\mathbf{L}(z_i)
p(z_i|\vy_i,\bar{\vc}_0,\mathbf{V}_0)
\dd z_i
\label{eq:F:def}
,
\\
\vg
&=&
\sum_{i=1}^{N_v}
\int_{z_i}
\mathbf{L}^T(z_i)
\mathbf{\Sigma}_i^{-1}(z_i,\mathbf{V}_0)
\vy_i
p(z_i|\vy_i,\bar{\vc}_0,\mathbf{V}_0)
\dd z_i
\label{eq:g:def}
,
\end{eqnarray}
subscript ``0'' indicates the previous value,
$\Sigma_i(z_i,\mathbf{V})
=
\mathbf{L}(z_i) \mathbf{V} \mathbf{L}^T(z_i) + \mathbf{Q}$
\cite[Eq.~7]{YiliZhengQiuWangDoerschukJOSA2012},
$p(z_i|\vy_i,\bar{\vc},\mathbf{V})$ is computed from
$p(\vy_i|z_i,\bar{\vc},\mathbf{V})
=
N(\vmu_i(z_i,\bar{\vc}),\Sigma_i(z_i,\mathbf{V}))(\vy_i)
$
\cite[Eq.~8]{YiliZhengQiuWangDoerschukJOSA2012},
and
$\vmu_i(z_i,\bar{\vc})=\mathbf{L}(z_i)\bar{\vc}$
\cite[Eq.~5]{YiliZhengQiuWangDoerschukJOSA2012}.
The update equation for the new value of $\mathbf{V}$ is to solve the
maximization problem~\cite[Eq.~35]{YiliZhengQiuWangDoerschukJOSA2012}
$\mathbf{V}^{\rm new}=\argmax_{\mathbf{V}} Q(\mathbf{V})$ where
\begin{eqnarray}
Q(\mathbf{V})
&=&
\sum_{i=1}^{N_v}
\int_{z_i}
\ln
\det(
\mathbf{\Sigma}^{-1}(z_i,\mathbf{V}_0)
)
p(z_i|\vy_i,\bar{\vc}_0,\mathbf{V}_0)
\dd z_i
\nonumber\\
-
\sum_{i=1}^{N_v}
&& \!\!\!\!\!\!\!\!
\int_{z_i}
\!\!\!
\tr[
\mathbf{\Sigma}^{-1}(z_i,\mathbf{V}_0)
\mathbf{N}_i(\vy_i,z_i,\bar{\vc}_0)
]
p(z_i|\vy_i,\bar{\vc}_0,\mathbf{V}_0)
\dd z_i
\label{eq:QforVoptimization:def}
\end{eqnarray}
and
$\mathbf{N}_i(\vy_i,z_i,\bar{\vc})=(\vy_i-\vmu_i(z_i,\bar{\vc}))
(\vy_i-\vmu_i(z_i,\bar{\vc}))^T$~\cite[Eq.~34]{YiliZhengQiuWangDoerschukJOSA2012},
which is solved by
MATLAB's {\tt fmincon} (option ``trust-region-reflective'') with symbolic
cost, gradient of the cost, and Hessian of the cost.
Any set of basis functions $F_\omega(\vx)$ could be used within
$\mathbf{L}(z_i)$ and the only constraint is that $\mathbf{V}$ is positive
definite.
\par
The estimator of Ref.~\cite{YiliZhengQiuWangDoerschukJOSA2012} has two
capabilities that are not described in this paper:
(i)~If the collection of cryo-EM images are unlabeled images from a known
number of classes, the {\em a~priori} pmf for class membership is estimated
and a separate value of $\bar{\vc}$ and $\mathbf{V}$ is estimated for each
class.
(ii)~If desired, the uniform {\em a~priori} pdf/pmf on the nuisance
parameters, such as the projection orientations, can be replaced by an
unknown pdf/pmf that is estimated.
The treatment of symmetry described in this paper preserves these
capabilities but they are not described in this paper both in order to
simplify notation and because the examples of Section~\ref{sec:results} do
not require these capabilities.
\section{Symmetric statistics}
\label{sec:symmetrysymmetricstatistics}
The relevant symmetries are the 3-D rotational point group symmetries,
which are the icosahedral $I$, octahedral $O$, tetrahedral $T$, cyclic
$C_n$ ($n\in\{1,2,\dots\}$), and dihedral $D_n$ ($n\in\{1,2,\dots\}$)
groups.
Once a coordinate system is chosen, each of these groups can be described
as a collection of matrices, denoted by $R_\beta\in\mathbb{R}^{3\times 3}$
($\beta\in\{1,\dots,\Ngroup\}$), which are rotation matrices, i.e.,
$R_\beta^{-1}=R_\beta^T$ and $\det R_\beta=+1$.
The value of $\Ngroup$ is 60, 24, 12, $n$, and $2n$ for $I$, $O$, $T$, $C_n$,
and $D_n$, respectively.
\par
The symmetry can be applied in three different ways.
(i)~Assume that the particles in a class are identical and that the common
electron scattering intensity, which is denoted by $\rho(\vx)$, has the
symmetry,
i.e., $\rho(R_\beta^{-1}\vx)=\rho(\vx)$ for all
$\beta\in\{1,\dots,\Ngroup\}$ and $\vx\in\mathbb{R}^3$.
Algorithms and software for computing structures in this situation are
widely available (Section~\ref{sec:introduction:heterogeneity}).
(ii)~Assume that each particle in a class is different but that each
particle has the symmetry so that the electron scattering intensity of the
$i$th particle, which is denoted by $\rho_i(\vx)$, satisfies
$\rho_i(R_\beta^{-1}\vx)=\rho_i(\vx)$ for all $\beta\in\{1,\dots,\Ngroup\}$
and $\vx\in\mathbb{R}^3$.
Algorithms and software for computing structures in this situation are
available (Section~\ref{sec:introduction:heterogeneity}), but produce
anomalous results when symmetry is
present~\cite[p.~173]{Ludtke.MethodsEnzymology.2016}.
(iii)~Assume that each particle in a class is different, that no particle has
the symmetry, but the statistics of the i.i.d.\ ensemble of particles has
the symmetry as is described in the following paragraph.
The authors are unaware of prior work of this type.
\par
One approach to introduce symmetric statistics is to require symmetry in
all of the finite-dimensional probability measures that together, via
Kolmogorov's extension theorem~\cite[Theorem~2.1.5 p.~11]{Oksendal2003},
define the electron scattering intensity stochastic process.
If the probability measures can be described by pdfs, then the symmetry
condition is
$p_{R_\beta^{-1}\vx_1,\dots,R_\beta^{-1}\vx_k}(\rho_1,\dots,\rho_k)
=
p_{\vx_1,\dots,\vx_k}(\rho_1,\dots,\rho_k)$
for all $\beta\in\{1,\dots,\Ngroup\}$, $k\in\{1,2,\dots\}$,
$\vx_1,\dots,\vx_k\in\mathbb{R}^3$, and $\rho_1,\dots,\rho_k\in\mathbb{R}$.
Define the mean function and covariance function of the electron scattering
intensity, denoted by $\bar\rho(\vx)$ and $\CrhoREAL(\vx_1,\vx_2)$,
respectively, by $\bar\rho(\vx)=\expect[\rho(\vx)]$ and
$\CrhoREAL(\vx_1,\vx_2)=\expect[(\rho(\vx_1)-\bar\rho(\vx_1))(\rho(\vx_2)-\bar\rho(\vx_2))]$,
respectively.
From the $k=1$ and $k=2$ instances of the symmetry condition for the pdfs,
it is straightforward to compute that for all
$\vx,\vx_1,\vx_2\in\mathbb{R}^{3}$ and $\beta\in\{1,\dots,\Ngroup\}$,
\begin{eqnarray}
\bar\rho(R_\beta^{-1} \vx)
&=&
\bar\rho(\vx)
\label{eq:symmetry:firstorder}
\\
\CrhoREAL(R_\beta^{-1} \vx_1, R_\beta^{-1} \vx_2)
&=&
\CrhoREAL(\vx_1,\vx_2)
\label{eq:symmetry:secondorder}
.
\end{eqnarray}
If the probability distribution for the electron scattering intensity is
Gaussian then
(\ref{eq:symmetry:firstorder})--(\ref{eq:symmetry:secondorder})
imply that the
symmetry condition for the pdfs is satisfied for all values of $k$.
\section{Description of $\rho(\vx)$: Basis functions}
\label{sec:BasisFunction}
In order to impose the symmetry conditions of
(\ref{eq:symmetry:firstorder}) and (\ref{eq:symmetry:secondorder}) on the
electron scattering intensity $\rho(\vx)$ when $\rho(\vx)$ is described by
(\ref{eq:rho:general}), it is helpful for the basis functions
$F_\omega(\vx)$ to have particular properties under rotations which are
described in this section.
As described in Section~\ref{sec:combiningsymmetryandheterogeneity}, in the
remainder of the paper we focus on the icosahedral $I$ group because
there exist real irreducible representations (irreps) of the group, the
natural coordinate system is spherical coordinates, $I$ is important in the
structural biology of viruses, and $I$ is the group in our example.
The octahedral $O$ group has parallels with $I$, there exist real irreps
and the natural coordinate system is spherical coordinates, so the the
discussion of $I$ can be carried over to $O$ with no significant changes.
The tetrahedral $T$, cyclic $C_n$, and dihedral $D_n$ groups require
complex irreps and/or cylindrical coordinates and the necessary results are
outlined in
Appendices~\ref{sec:constraintsonthemomentsoftheweights:complex}
and~\ref{sec:nonsphericalcoordinatesystems}.
\begin{figure}
	\begin{center}
		\begin{tabular}{ccc}
			\includegraphics[width=2.4cm]{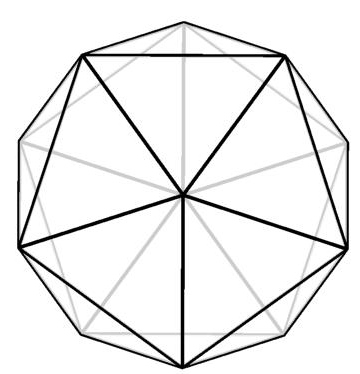}
			&
			\includegraphics[width=2.4cm]{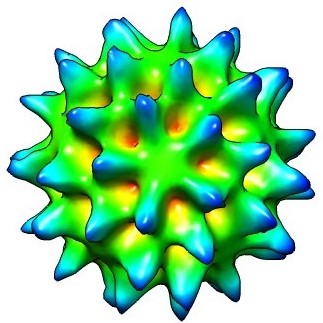}
			&
			\includegraphics[width=2.4cm]{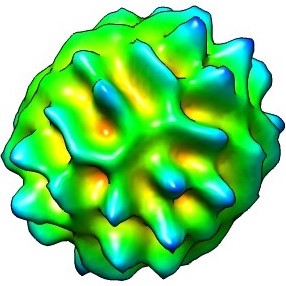}
			\\
			&
			$p=1$, $j=1$,
			&
			$p=2$, $j=1$,
			\\
			&
			$n=1$
			&
			$n=1$
			\\
			\includegraphics[width=2.4cm]{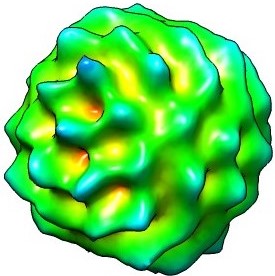}
			&
			\includegraphics[width=2.4cm]{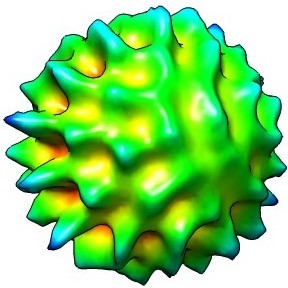}
			&
			\includegraphics[width=2.4cm]{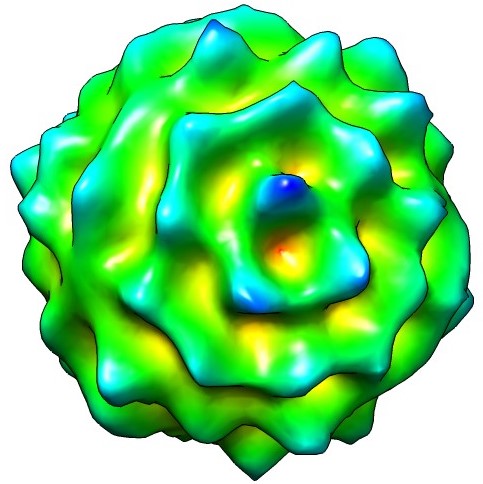}
			\\
			$p=3$, $j=1$,
			&
			$p=4$, $j=1$,
			&
			$p=5$, $j=1$,
			\\
			$n=1$
			&
			$n=2$
			&
			$n=2$
		\end{tabular}
	\end{center}
	\vspace*{-.1in}
	\caption{
		\label{fig:irredrep}
		An icosahedron with a 5-fold axis pointing out of the page and example
		angular basis functions in the same orientation with $l=15$ and
		$p\in\{1,\dots,\Nirrep\}$.
		The surfaces of 3-D objects defined by $\xi(\vx)=1$ for $x\le
		\kappa_1+\kappa_2 I_{p,j;l,n}(\vx/x)$ and $\xi(\vx)=0$ otherwise, where $\kappa_1$
		and $\kappa_2$ are chosen so that
		$\kappa_1+\kappa_2 I_{p,j;l,n}(\vx/x)$ varies between 0.5 and 1.0,
		are visualized by UCSF
		Chimera~\cite{PettersenHuangCouchGreenblattMengFerrin2004} where the color
		indicates the distance from the center of the object.
		Let $N_E(l)$ denote the number of extreme values (maxima and minima) of
		$I_{p,j;l,n}(\vx/x)$ as $\vx$ varies around any great circle grows.
		The function $N_E(l)$ is a linear function of $l$ with a positive slope.
		Because $I_{p,j;l,n}(\vx/x)$ can be negative, zero, or positive,
		visualizing $I_{p,j;l,n}(\vx/x)$ in the style of this figure requires some
		transformation of $I_{p,j;l,n}(\vx/x)$ and the affine transformation
		defined by $\kappa_1$ and $\kappa_2$ is simple.
	}
\end{figure}
\par
Using spherical coordinates, we assume that
$F_\omega(\vx)=I_\omega(\vx/x)\psi_\omega(x)$ where $I_\omega(\vx/x)$ is
the angular basis function and is a linear combination of spherical
harmonics\footnote{Spherical harmonics are denoted by
  $Y_{l,m}(\theta,\phi)$ where the degree $l$ satisfies $l\in\mathbb{N}$,
  the order $m$ satisfies $m\in\{-l,\dots,l\}$ and $(\theta,\phi)$ are the
  angles of spherical coordinates with $0\leq \theta\leq \pi$ and
  $0\leq\phi\leq 2\pi$~\cite[Section~14.30, pp.~378--379]{OlverLozierBoisvertClark2010}.}
and $\psi_\omega(x)$ is the radial basis function and is a linear
combination of spherical Bessel functions as was used
previously~\cite{DoerschukJohnsonIT2000}.
The properties of $F_\omega(\vx)$ under rotational symmetries involves only
$I_\omega(\vx/x)$ so $I_\omega(\vx/x)$ is the focus of this section.
\par
Ref.~\cite{XuDoerschukBasisFunctions2017} derives a basis $I_\omega(\vx/x)$
that has particular properties under the operations of a rotational point
group.
In order to describe the properties of the basis, it is necessary to
describe the idea of representations.
A representation (rep) of a rotational point group, which is a finite
group, is a set of matrices that are homomorphic under matrix
multiplication to the group elements~\cite[p.~61]{Miller1972}.
The irreducible representations (irreps) of a finite group are a set of
unitary reps such that all unitary reps can be decomposed into a direct sum
of the irreps~\cite[Theorem~3.3 p.~69]{Miller1972}.
As is described in the first paragraph of this section, for $I$ and $O$,
all irreps can be chosen to be real-valued orthonormal rather than
complex-valued unitary and this is the situation considered in the
remainder of this paper except for
Appendices~\ref{sec:constraintsonthemomentsoftheweights:complex}
and~\ref{sec:nonsphericalcoordinatesystems}.
Let $\Nirrep$ be the number of irreps of the $\Ngroup$-order group $G$.
Let the set of real-valued matrices in the $p$th irrep be denoted by
$\Gamma^p(g)\in\mathbb{R}^{d_p\times d_p}$ for all $g\in G$ where
$p\in\{1,\dots,\Nirrep\}$.
For the icosahedral group, $\Nirrep=5$, $\Ngroup=60$, and
$d_p=1,3,3,4,5$ for $p=1,2,3,4,5$, respectively.
The vector-valued basis functions $I_{p,\zeta}(\vx/x)$
($\omega=(p,\zeta)$) were derived~\cite{XuDoerschukBasisFunctions2017} by
applying real-valued generalized projection operators, which were
constructed from real-valued irreps, to real-valued spherical harmonics.
This results in the index $\zeta$ being the ordered pair $(l,n)$,
where $l$ indexes the subspace defined by spherical harmonics of degree $l$,
and $n$ indexes the different basis functions within the subspace.
The basis functions $I_{p;l,n}(\vx/x)$ have the properties
\begin{enumerate}
\item
Each $I_{p;l,n}$ is a $d_p$-dimensional real-valued vector function, i.e.,
$I_{p;l,n}\in \mathbb{R}^{d_p}$.
\item
The $I_{p;l,n}$ functions are orthonormal on the surface of the sphere,
i.e.,
\begin{equation}
\int
I_{p;l,n}(\vx/x)
I_{\pp;\lp,\np}^T(\vx/x)
\dd\Omega
=
\delta_{p,\pp}
\delta_{l,\lp}
\delta_{n,\np}
I_{d_p}
\label{eq:orthonormalityofangularbasisfunctions}
\end{equation}
\item
The subspace of square integrable functions on the surface of the sphere
defined by spherical harmonics of degree $l$, contains a set of $I_{p;l,n}$
functions with a total of $2l+1$ components.
\item
Each $I_{p;l,n}$ function has a specific transformation property under
rotations from the rotational point group~\cite[p.~20]{Cornwell1984}, in
particular,
\begin{equation}
I_{p;l,n}(R_g^{-1} \vx/x)
=
(\Gamma^p(g))^T
I_{p;l,n}(\vx/x)
\label{eq:transformationEqn}
\end{equation}
where ${}^T$ is transpose and $R_g$ is the
$3\times 3$ rotation matrix corresponding to the $g$th element of the
group.
\end{enumerate}
\par
Examples of $I_{p,j;l,n}(\vx/x)$ functions for $l=15$, where
$I_{p,j;l,n}(\vx/x)$ is the $j$th component of the $d_p$-dimensional
vector-valued function $I_{p;l,n}(\vx/x)$, are visualized in
Figure~\ref{fig:irredrep}.
Note, as expected, that the $I_{p=1,j;l,n}(\vx/x)$ function exhibits all of
the symmetries of an icosahedron since the $p=1$ irrep is the identity
irrep for which $\Gamma^{p=1}(g)=1$ for all $g\in G$.
\section{Symmetric statistics requires constraints on the moments of the weights}
\label{sec:constraintsonthemomentsoftheweights}
We use the basis functions of Section~\ref{sec:BasisFunction} in
(\ref{eq:rho:general}).
Since $\rho(\vx)$ is real valued and the basis functions are real valued,
it is only necessary to consider real-valued weights.
We include the radial dependence of the basis function in the weight (which
is a vector since the angular basis functions are vector valued) which
leads to writing (\ref{eq:rho:general}) in the form
\begin{equation}
\rho(\vx)
=
\sum_{p=1}^{\Ngroup}
\sum_{l=0}^\infty
\sum_{n=0}^{N_n(p,l)-1}
c_{p,\zeta=(l,n)}^T(x)
I_{p,\zeta=(l,n)}(\vx/x)
\label{eq:I2rho}
\end{equation}
where
$I_{p,\zeta}(\vx/x)\in\mathbb{R}^{d_p}$ as in Section~\ref{sec:BasisFunction}
($p$ indexes the irrep, $\zeta$ is a shorthand for the $l$ and $n$ indices,
and $d_p$ is the dimension of the $p$th irrep),
$c_{p,\zeta}(x)\in\mathbb{R}^{d_p}$,
and
the integer $N_n(p,l)$ (which is the number of basis functions of index $l$
that transform as irrep $p$) is computed during the computation of
$I_{p,\zeta}(\vx/x)$.
In more detail, the abstract index $\omega$ of (\ref{eq:rho:general})
becomes the triple $(p,l,n)$ and the product $c_\omega F_\omega(\vx)$
(weight times basis function) of (\ref{eq:rho:general}) becomes the product 
$c_{p,\zeta=(l,n)}^T(x) I_{p,\zeta=(l,n)}(\vx/x)$.
In (\ref{eq:radialexpansion}), $c_{p,\zeta=(l,n)}(x)$ is described as a
weighted sum of basis functions with weights denoted by $c_{p,\zeta,q}$ and
basis functions denoted by $\psi_{p,\zeta,q}(x)$,
i.e.,
$c_{p,\zeta}(x)
=
\sum_{q=1}^{N_q}
c_{p,\zeta,q}
\psi_{p,\zeta,q}(x)$.
Then $c_\omega$ of (\ref{eq:rho:general}) becomes $c_{p,\zeta,q}$ and
$F_\omega(\vx)$ of (\ref{eq:rho:general}) becomes $\psi_{p,\zeta,q}(x)
I_{p,\zeta=(l,n)}(\vx/x)$.
\par
The constraints on the mean function and covariance function of
$c_{p,\zeta}(x)$ are derived in Appendices~\ref{sec:mean:derivation}
and~\ref{sec:covariance:derivation} and the results are stated in this
paragraph.
The mean obeys the constraint
\begin{equation}
\bar c_{p,\zeta}(x)
=
\left\{
\begin{array}{ll}
\mbox{arbitrary}
,
&
p\in\{1\}
\\
0_{d_p,1}
,
&
p\in\{2,\dots,\Nirrep\}
\end{array}
\right.
\label{eq:barc:constraint:xdependent}
.
\end{equation}
The covariance obeys the constraint
\begin{equation}
\CcREAL(x_1,x_2)
=
\left\{
\begin{array}{ll}
c_{p_1}(\zeta_1,x_1;\zeta_2,x_2)
I_{d_{p_1}}
,
&
p_1=p_2
\\
0_{d_{p_1},d_{p_2}}
,
&
\mbox{otherwise}
\end{array}
\right.
\label{eq:Cc:constraint:xdependent}
\end{equation}
where $c_{p_1}(\zeta_1,x_1;\zeta_2,x_2)\in\mathbb{C}$.
\par
The results for the mean function and covariance function of
$c_{p,\zeta}(x)\in\mathbb{R}^{d_p}$ given in
(\ref{eq:barc:constraint:xdependent}) and~(\ref{eq:Cc:constraint:xdependent})
are results concerning functions.
To reduce these to results concerning finite-dimensional vectors for
computation, we assume
that $c_{p,\zeta}(x)\in\mathbb{R}^{d_p}$ is described as a linear combination
of scalar real-valued orthonormal basis functions with vector real-valued
weights.
The scalar real-valued orthonormal basis functions are the radial basis
functions, denoted by $\psi_\omega(x)$, of Section~\ref{sec:BasisFunction}.
An exact description would require an infinite number of basis functions
but we truncate the number of basis functions to $N_q$, which sets a limit on
the spatial resolution that can be achieved.
Based on detailed notation and derivations in
Appendix~\ref{sec:finitedimensional},
the results are
\begin{eqnarray}
\bar c_{p,\zeta,q}
\!\!
&=&
\!\!
\left\{
\begin{array}{ll}
\mbox{arbitrary}, & p=1 \\
0_{d_p,1}, & p\in\{2,\dots,\Nirrep\}
\end{array}
\right.
\label{eq:barc:constraint:xindependent}
\\
\mathbf{V}_{p_1,\zeta_1,q_1;p_2,\zeta_2,q_2}
\!\!
&=&
\!\!
\left\{
\begin{array}{ll}
v_{p_1}(\zeta_1,q_1;\zeta_2,q_2)
I_{d_{p_1}}
,
&
p_1=p_2
\\
0_{d_{p_1},d_{p_2}}
,
&
\mbox{otherwise}
\end{array}
\right.
\label{eq:Cc:constraint:xindependent}
\end{eqnarray}
where $v_{p_1}(\zeta_1,q_1;\zeta_2,q_2)$ is arbitrary.
The goal of this paper is to estimate $\bar c_{p,\zeta,q}$ and
$\mathbf{V}_{p_1,\zeta_1,q_1;p_2,\zeta_2,q_2}$ or equivalently
$v_{p_1}(\zeta_1,q_1;\zeta_2,q_2)$ from image data.
\par
Equation (\ref{eq:Cc:constraint:xindependent})
implies structure for the complete
covariance matrix $\mathbf{V}$.
Suppose that the basis functions are enumerated with indices changing in
the order $p$ (slowest), $\zeta$, $q$, and $j$ (fastest) where $j$ is the
index that enumerates the elements of the vector $I_{p,\zeta}(\vx/x)$.
Then the $\mathbf{V}$ matrix is a $\Nirrep\times\Nirrep$ block matrix where the size
of the blocks is determined by the values of the $p$ indices and only the
diagonal blocks are nonzero (due to the ``$p_1=p_2$'' condition in
(\ref{eq:Cc:constraint:xindependent})).
Furthermore, each block is constructed of $N_{\zeta,q}\times N_{\zeta,q}$
subblocks (where $N_{\zeta,q}$ is the total number of $(\zeta,q)$ pairs)
and each subblock is proportional to the $d_p\times d_p$ identity matrix
with proportionality constant $v_{p_1}(\zeta_1,q_1;\zeta_2,q_2)$.
\section{Maximum likelihood reconstruction with symmetric statistics}
\label{sec:MLE}
The estimator of Section~\ref{sec:MLreconstruction} can be
applied to any set of basis functions $F_\omega(\vx)$, has no constraints
on the mean of the coefficients $\bar{\vc}$, and has only a
positive-definite constraint on the covariance of the coefficients
$\mathbf{V}$.
In this section, we specialize to the basis functions of
Section~\ref{sec:BasisFunction} and add the constraints of
Section~\ref{sec:constraintsonthemomentsoftheweights} on the statistics
$\bar{\vc}$ and $\mathbf{V}$ of the coefficients.
\par
In order to incorporate the basis functions of
Section~\ref{sec:BasisFunction} and the constraints into the estimator of
Section~\ref{sec:MLreconstruction},
we think of the mean vector $\bar{\vc}$ and covariance matrix $\mathbf{V}$
as functions of a further level of
parameterization~\cite{XuDoerschukEMBC2016}.
In particular:
(i) The mean is $\bar{\vc}(\mu)$ where $\mu$ is the arbitrary mean for the
$p=1$ basis functions since the $p\in\{2,\dots,\Ngroup\}$ basis functions
all have zero mean (see (\ref{eq:barc:constraint:xindependent})).
(ii) The covariance is $\mathbf{V}(\vv)$ where $\vv$ is composed of the values
$v_{p_1}(\zeta_1,q_1;\zeta_2,q_2)$ that are arbitrary except for the
requirement that $\mathbf{V}(\vv)$ be positive definite
(see (\ref{eq:Cc:constraint:xindependent})).
Using the additional level of parameterization, the gradient and Hessian
can be computed via the chain rule.
These equations compute the necessary gradient and Hessian in terms of
larger vectors and matrices that are then reduced in size.
While this approach fits well within the MATLAB software used in
Refs.~\cite{YiliZhengQiuWangDoerschukJOSA2012,QiuWangMatsuiDomitrovicYiliZhengDoerschukJohnsonJSB2012,TangKearneyQiuWangDoerschukBakerJohnsonJMolRecog2014,DomitrovicMovahedBothnerMatsuiQiuWangDoerschukJohnsonJMB2013,YunyeGongVeeslerDoerschukJohnsonJSB2016,DoerschukGongXuDomitrovicJohnsonCurOpinVirology2016},
more efficient approaches may be possible.
\par
We use ``\hetero'' to indicate the ideas, algorithms, and software of this
paper.
They are a generalization of Ref.~\cite{YiliZhengQiuWangDoerschukJOSA2012}
so we use ``{\heterosymparticles}'' to indicate
Ref.~\cite{YiliZhengQiuWangDoerschukJOSA2012} and
``{\heterosymstatistics}'' to indicate the generalization.
In ``{\heterosymparticles}'' the particles are heterogeneous and each
particle has the symmetry (use $p=1$ only)
while in ``{\heterosymstatistics}'' the particles
are heterogeneous, no particle has the symmetry, but the statistics of the
particles have the symmetry (use $p\in\{1,\dots,\Nirrep\}$).
Absence of symmetry is a symmetry with just one symmetry operator, the
identity operator, and the ideas, algorithms, and software include this
case, which we denote by ``{\heteroasymmetry},'' since the {\heterosymparticles}
and {\heterosymstatistics} ideas are identical in the asymmetric case.
\par
\begin{figure}
\begin{center}
\includegraphics[width=3.0in]{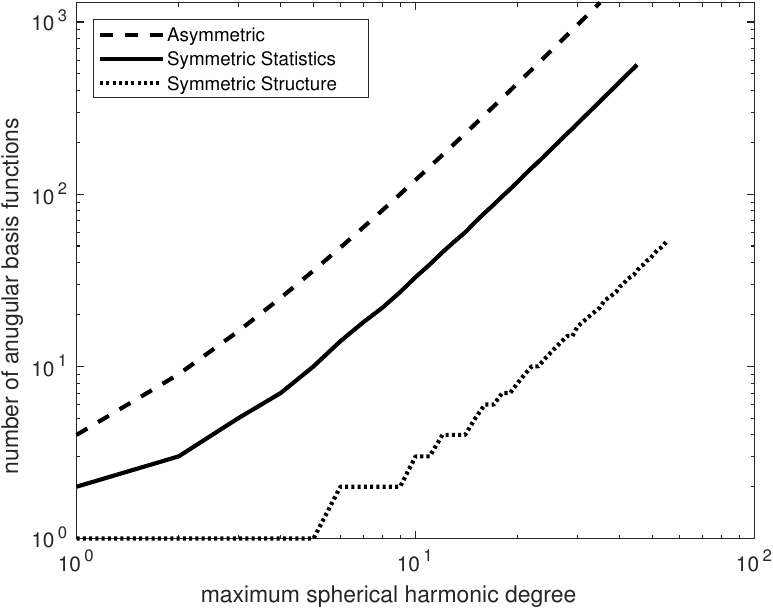}
\end{center}
\vspace*{-.2in}
\caption{
\label{fig:computationalcomplexity}
The number of variance parameters as a function of the maximum degree
$l_{\rm max}$ for the three different estimators where two of the
estimators incorporate icosahedral symmetry.
}
\end{figure}

The scale of computing for estimators using the three different symmetry
assumptions is described in this paragraph
in terms of the number of parameters that must be estimated from the data.
Figure~\ref{fig:computationalcomplexity} shows the number of variance
parameters that must be estimated (diagonal covariance matrix) as a
function of the upper limit on the degree $l$ of the spherical harmonics
$Y_{l,m}(\theta,\phi)$ that are included in the mathematical model of the
electron scattering intensity.
This upper limit is denoted by $l_{\rm max}$.
The number of variance parameters is shown for three cases:
(i)~{\heteroasymmetry}, 
(ii)~{\heterosymparticles}, 
and
(iii)~{\heterosymstatistics}. 
For both {\heterosymparticles} and {\heterosymstatistics}, the number of
mean parameters is equal to the number of variance parameters for
{\heterosymparticles} while for {\heteroasymmetry} the number of mean
parameters equals the number of variance parameters.
The curve for the {\heteroasymmetry} 
case is exactly $(l_{\rm max}+1)^2$ and the other two curves are also
approximately quadratic in $l_{\rm max}$.
The attractive aspect of the computational complexity of the
{\heterosymstatistics} approach is that the {\heterosymstatistics} curve is
substantially below the {\heteroasymmetry} 
curve even though every electron scattering intensity in the
{\heterosymstatistics} approach is asymmetrical.
\par
In this paragraph we relate the isotropic resolution goal for the
mathematical model, denoted by a distance $\gamma$, to
$l_{\rm max}$ and
$N_q$ (the number of radial basis functions $\psi_{p,\zeta,q}(x)$ (see
(\ref{eq:radialexpansion})))
by counting zero crossings
of $I_{p,\zeta=(l,n)}(\vx/x)$ for $l_{\rm max}$ and
of $c_{p,\zeta=(l,n)}(x)$ for $N_q$.
The particle has radius $R$.
In order to achieve resolution $\gamma$ in the radial direction it is
certainly necessary to have $R/\gamma$ zero crossings of
$c_{p,\zeta=(l,n)}(x)$ between $x=0$ and $x=R$ and therefore
$N_q=R/\gamma$ is required.
In order to achieve resolution $\gamma$ in the angular directions it is
certainly necessary to have $2\pi R/\gamma$ zero crossings of
$I_{p,\zeta=(l,n)}(\vx/x)$ around any great circle and therefore
$l_{\rm max}=2\pi R/\gamma$ is required.
The total number of parameters is the number of mean parameters plus the
number of variance parameters.
For the three different symmetry assumptions described in the previous
paragraph, the number of parameters are as follows:
{\heteroasymmetry} requires $(R/\gamma)2\calK_{\mbox{Asym}}(2\pi R/\gamma)$,
{\heterosymparticles} requires $(R/\gamma)2\calK_{\mbox{SymPart}}(2\pi R/\gamma)$,
and
{\heterosymstatistics} requires
$(R/\gamma)[\calK_{\mbox{SymStat}}(2\pi R/\gamma)+\calK_{\mbox{SymPart}}(2\pi R/\gamma)]$.
The values of 
$\calK_{\mbox{Asym}}(\cdot)$,
$\calK_{\mbox{SymStat}}(\cdot)$,
and
$\calK_{\mbox{SymPart}}(\cdot)$
can be read from Figure~\ref{fig:computationalcomplexity}.
\par
The algorithm has the structure of nested iterations where the inner
loop is within MATLAB's {\tt fmincon}, used to update $\mathbf{V}$,
and the outer loop is the E-M iteration loop.
The computational complexity of the algorithm depends on proprietary
behavior of {\tt fmincon}, iteration limits, and convergence criteria and
so is difficult to quantify.
Typically, 8 iterations are used in the inner {\tt fmincon} loop.
Typically 20--200 iterations are used in the outer E-M loop depending on
the quality of the initial condition.
Calculation of $L(z_i)$ for the various values of $z_i$ in the quadrature
rule that approximates the integrals in
(\ref{eq:F:def})--(\ref{eq:QforVoptimization:def})
is the dominant computational cost.
The computational complexity of this cost depends on proprietary behavior
of MATLAB's {\tt legendre} function and possibly other functions and so
is difficult to quantify.
Typically a rule with 5000 abscissas is used.
The computational complexity also depends on the number of pixels in an
image versus the number of basis functions retained in the truncated
orthonormal expansion (i.e., the dimension of
$\bar{\vc}$)~\cite[Section~6]{YiliZhengQiuWangDoerschukJOSA2012} since
these are the dimensions of the matrix $L$ which occurs throughout the
formulas of Section~\ref{sec:MLreconstruction}.
Considering just terms in the cost due to the number of images for one
iteration of either the inner or the outer loop, these terms are linear
in the number of images.
\par
The calculations are performed in three steps.
(i) An initial spherically-symmetric reconstruction is estimated from the
set of images by using only $l=0$ basis functions which can be done by
least squares.
(ii) A homogeneous reconstruction, i.e., a reconstruction in which all
particles are assumed to be identical and have icosahedral symmetry, is
performed.
In terms of Section~\ref{sec:MLE}, this is the case where $p=1$,
$\vc_i=\bar\vc$, and $\mathbf{V}=0_{N_c,N_c}$ so this algorithm is denoted by
{\heterohomo}.
(iii) A heterogeneous reconstruction, i.e., a reconstruction in which no
particles are the same, is performed using the homogeneous reconstruction
$\bar\vc$ as an initial condition.
Two types of heterogeneous reconstruction are computed and contrasted.
One type of reconstruction uses only the $p=1$ basis functions
({\heterosymparticles}), so that each instance of the particle has the
icosahedral symmetry.
The other type of reconstruction uses the full set of
$p\in\{1,\dots,\Nirrep=5\}$ basis functions with the constraints of
Section~\ref{sec:constraintsonthemomentsoftheweights}
({\heterosymstatistics}), so that no instance of the particle has the
icosahedral symmetry but the first and second order statistics of the
electron scattering intensity $\rho(\vx)$ do have the icosahedral
symmetry.
The purpose of Steps~(i) and~(ii) is to provide an initial condition for
Step~(iii).
Various versions of Steps~(i) and~(ii) have been used, e.g., switch from
homogeneous to heterogeneous reconstructions at low resolution and then
increase the resolution, without interpretable differences in the final
reconstruction but we do not have a theory or comprehensive results that
can be presented here.
The {\heterohomo} followed by {\heterosymparticles} reconstructions used
basis functions with $p=1$, $l\in\{0,1,\dots,55\}$, and
$q\in\{1,2,\dots,20\}$
($q$ indexes the radial basis functions $\psi_{p,\zeta,q}(x)$ (see
(\ref{eq:radialexpansion})),
1060 total basis
functions) while the {\heterohomo} followed by {\heterosymstatistics}
reconstructions used $p\in\{1,\dots,\Nirrep\}$, $l\in\{1,2,\dots,10\}$, and
$q\in\{1,2,\dots,20\}$
(2020 total basis functions) where only smaller $l$ values can be used
because of the substantially larger number of basis functions that occur
when $p\in\{1,\dots,\Nirrep\}$ versus $p=1$.
Limited by our computer system, only diagonal $\mathbf{V}$ matrices are
considered, which is the simplest case of utilizing all basis functions
while guaranteeing the assumption of symmetric statistics.
Even with a more powerful computer, estimation of full covariance matrices
at the scale of $2020\times 2020$ would probably require regularization,
perhaps a sparseness regularizer, and would probably not be done with a
maximum likelihood
estimator~\cite{WainwrightTibshiraniHastieStatisticalLearningWithSparsity2015}.
\par
The mean of the electron scattering intensity,
$\bar\rho(\vx)=E[\rho(\vx)]$, is related to $\bar{\vc}$ by
(\ref{eq:barrho}) and (\ref{eq:meanofradialexpansion})
(or Ref.~\cite[Eq.~16]{YiliZhengQiuWangDoerschukJOSA2012}).
In the simpler notation of (\ref{eq:rho:general}), the result is
$\bar\rho(\vx)=\sum_{\omega\in\Omega} \bar{\vc}_\omega F_\omega(\vx)$.
The covariance of the electron scattering intensity,
$\CrhoREAL(\vx_1,\vx_2)=E[(\rho(\vx_1)-\bar\rho(\vx_1))(\rho(\vx_2)-\bar\rho(\vx_2))]$,
is related to $\mathbf{V}$ by
(\ref{eq:Crho}) and (\ref{eq:covarianceofradialexpansion})
(or Ref.~\cite[Eq.~18]{YiliZhengQiuWangDoerschukJOSA2012}).
In the simpler notation of (\ref{eq:rho:general}), the result is
$\CrhoREAL(\vx_1,\vx_2)=\sum_{\omega\in\Omega}
\sum_{\omega^\prime\in\Omega^\prime}
\mathbf{V}_{\omega,\omega^\prime}
F_\omega(\vx_1)
F_{\omega^\prime}(\vx_2)$.
The MLE provides estimates of $\bar{\vc}$ and $\mathbf{V}$.
Using these estimates in the place of the true values gives the estimates
of $\bar\rho(\vx)$ and $\CrhoREAL(\vx_1,\vx_2)$ that are used in this paper
and these estimates are themselves maximum likelihood
estimates~\cite[Thm.~7.2.10 p.~320]{CasellaBerger2002}.
We often visualize the estimate of the standard deviation function
$s(\vx)=\sqrt{\CrhoREAL(\vx,\vx)}$, which has the same
units as $\bar\rho(\vx)$ and is only 3-D instead of 6-D.
\par
All computations described in this paper were performed on a PC with two
Intel Xeon microprocessors (E5-2670, 2.60~GHz) each with 8 cores running
CentOS release 6.8 of GNU/Linux.
The software for this paper is about $10^4$ lines of MATLAB, using {\tt
  parfor} to achieve multi-core parallelism, running on 12 cores of the PC
where the limitation of 12 cores is set by our license not by the structure
of the parallelism.
Please contact PCD for a copy of the software.
\section{Quantification of performance}
A standard measure of resolution in cryo-EM is the Fourier shell
correlation (FSC) (Refs.~\cite[Eq.~2]{vanHeelUltramicroscopy1987},
\cite[Eq.~17]{HarauzvanHeelOptik1986},
\cite[p.~879]{BakerOlsonFullerMicrobiolMolBiolRev1999}) between the
estimated electron scattering intensities of two independent
reconstructions.
FSC is defined by
(Ref.~\cite[Eqs.~22--25]{YinZhengDoerschukNatarajanJohnsonJSB2003})
\begin{equation}
\mbox{FSC}(k)
=
\frac{
\int P_1(\vk) P_2^\ast(\vk) \dd\Omega^\prime
}{
\sqrt{
\left[\int |P_1(\vk)|^2 \dd\Omega^\prime\right]
\left[\int |P_2(\vk)|^2 \dd\Omega^\prime\right]
}
}
\label{eq:FSC:def}
\end{equation}
where $P_1(\vk)$ and $P_2(\vk)$ are the 3-D Fourier transforms of the
electron scattering intensities for the two structures and
$\vk=(k,\thetap,\phip)$ in spherical coordinates.
In the approach of this paper, the estimate of the electron scattering
intensity is the estimate of $\bar\rho(\vx)$.
The behavior of the FSC curve is influenced by the energy that
is defined by
\begin{equation}
E(k)
=
\int |P(\vk)|^2 \dd\Omega^\prime
\label{eq:Energy:def}
.
\end{equation}
The value of $k$ (denoted by $k_\ast$) at which the FSC curve first
decreases below a threshold is used to describe the level of similarity of
the two structures.
For instance, when the two structures are computed from disjoint sets of
images of the same particle, then the quantity $1/k_\ast$ is interpreted as an
isotropic spatial resolution of the computation.
In the calculations of this paper, the two sets of images are processed
completely independently.
In this paper, the threshold used is 1/2.
While it is standard in structural biology to apply FSC to determine the
resolution of the structure,
there is no standard for measuring resolution of the covariance function
$\CrhoREAL(\vx_1,\vx_2)$, variance function $v(\vx)$, or
standard deviation function $s(\vx)$.
In this paper, we will report norms of the difference of $\mathbf{V}$
values.
\section{Reconstruction Results}
\label{sec:results}
Numerical results based on cryo-EM images of two different particles, the
bacteriophage
HK97~\cite{HendrixJohnson.HK97assemblymaturation.ExpMedBio.2012} and the
virus N$\omega$V~\cite{MatsuiLanderKhayatJohnsonPNAS2010}, are described in this
section.
Numerical results on both synthetic and experimental images of a Virus Like
Particle (VLP) derived from bacteriophage
HK97~\cite{HendrixJohnson.HK97assemblymaturation.ExpMedBio.2012} are
described in detail in Sections~\ref{sec:results:simulated}
and~\ref{sec:results:HK97:experimental}, respectively.
The VLP is essentially the bacteriophage minus the bacteriophage's portal,
tail, and genome leaving only the icosahedrally symmetric capsid.
HK97 has a complicated lifecycle wherein it first self-assembles in near
equilibrium conditions and then undergoes a sequence of essentially
irreversible maturation transformations, which result in a robust particle
capable of surviving outside of the host cell.
One of the first steps of maturation is the digestion by a virally-encoded
protease of the so-called $\delta$ domain of the $60\times 7$ copies of the
capsid peptide that together make up the capsid of the bacteriophage.
(The $\delta$ domain is roughly the region between radii 93~{\AA} and
193~{\AA}~\cite{VeeslerKhayatKrishnamurthySnijderHuangHeckAnandJohnsonHK97Structure2014,YunyeGongVeeslerDoerschukJohnsonJSB2016}
where the outter radius of the particle is
254~{\AA}~\cite[PDB~3QPR]{HuangKhayatLeeGertsmanDudaHendrixJohnsonHK97JMB2011})
The experimental
images~\cite{VeeslerKhayatKrishnamurthySnijderHuangHeckAnandJohnsonHK97Structure2014}
come from the particle, denoted by
{\withpro}~\cite{YunyeGongVeeslerDoerschukJohnsonJSB2016}, that contains a
protease that is defective so that the particle is trapped in the Prohead~I
step of maturation.
The average outer radius of the capsid is
254~{\AA}~\cite[PDB~3QPR]{HuangKhayatLeeGertsmanDudaHendrixJohnsonHK97JMB2011}
and the sphere in which the reconstruction is computed has radius 280~{\AA}.
\par
In addition, numerical results on experimental images of the virus
N$\omega$V~\cite{MatsuiLanderKhayatJohnsonPNAS2010} are described in detail
in Section~\ref{sec:results:NwV:experimental}.
N$\omega$V has a complicated lifecycle wherein it self-assembles in near
equilibrium conditions at neutral pH, undergoes a large reduction in
diameter when exposed to acidic conditions, and finally undergoes $60\times
4$ self-catalyzed cleavage reactions in the peptide capsid which makes the
diameter reduction irreversible.
The experimental images~\cite{MatsuiLanderKhayatJohnsonPNAS2010} come from
particles 30~minutes after the pH change.
The average outer radius of the capsid is
211~{\AA}~\cite[PDB~1OHF]{HelgstrandMunchiJohnsonLiljasNwVVirology2004} and
the sphere in which the reconstruction is computed has radius 230~{\AA}.
\par
The mathematical model of Section~\ref{sec:MLreconstruction} has
parameters.
The radius $R_2$ of the ball in which the reconstruction is computed can be
estimated directly from micrographs.
The measurement noise variance $Q$ is estimated by the average over
particles of the sample variance in an annulus surrounding the particle.
As is described in Section~\ref{sec:MLE}, the truncation of the infinite
series of (\ref{eq:rho:general}) can be related to spatial resolution.
The truncation is described by $l_{\rm max}$ (the highest degree of the
spherical harmonics that are retained) and $N_q$ (the number of radial basis
functions $\psi_{p,\zeta,q}(x)$ (see (\ref{eq:radialexpansion})) that are
retained).
We have always retained all harmonics with $l\le l_{\rm max}$ and
$q\le N_q$ but that is not required.
While the formulas of Section~\ref{sec:MLE} allow selecting $l_{\rm max}$
and $N_q$ based on spatial resolution goals, we have generally been forced
to select based on limitations of our computer system
(Section~\ref{sec:MLE}).
\par
The optimization algorithm for computing the maximum likelihood estimator
has parameters.
We always start with a spherically-symmetric initial condition
($l_{\rm max}=0$) for which the estimator can be computed by a linear least
squares problem.
Spatial resolution is then increased stepwise.
We do not have a theory for whether it is better to first achieve high
resolution by {\heterosymparticles} and then switch to
{\heterosymstatistics} versus achieve high resolution directly with
{\heterosymstatistics} nor a theory for step sizes and are currently
experimenting with these parameters.
With the computer system described in Section~\ref{sec:MLE}, these choices
imply approximately one week calculations.
For each resolution step, we use the generalized expectation-maximization
algorithm (including the integration rule for the expectation and the
maximization algorithm for $V$) described in
Ref.~\cite{YiliZhengQiuWangDoerschukJOSA2012}.
\subsection{Simulated images: motivated by HK97}
\label{sec:results:simulated}
\begin{figure}[t]
\begin{tabular}{c@{\hspace{0.05in}}c@{\hspace{0.05in}}c}
\includegraphics[height=1.1in]{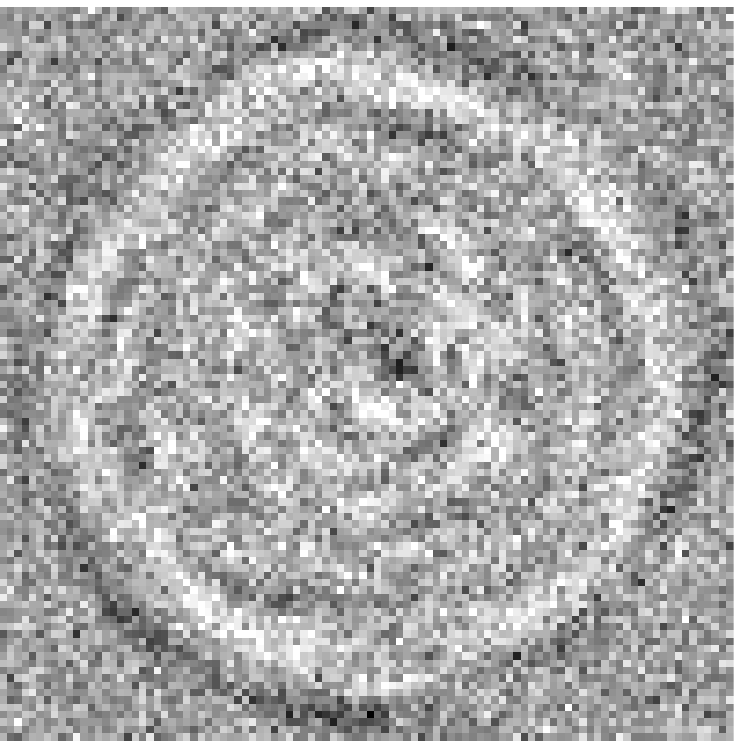}
&
\includegraphics[height=1.1in]{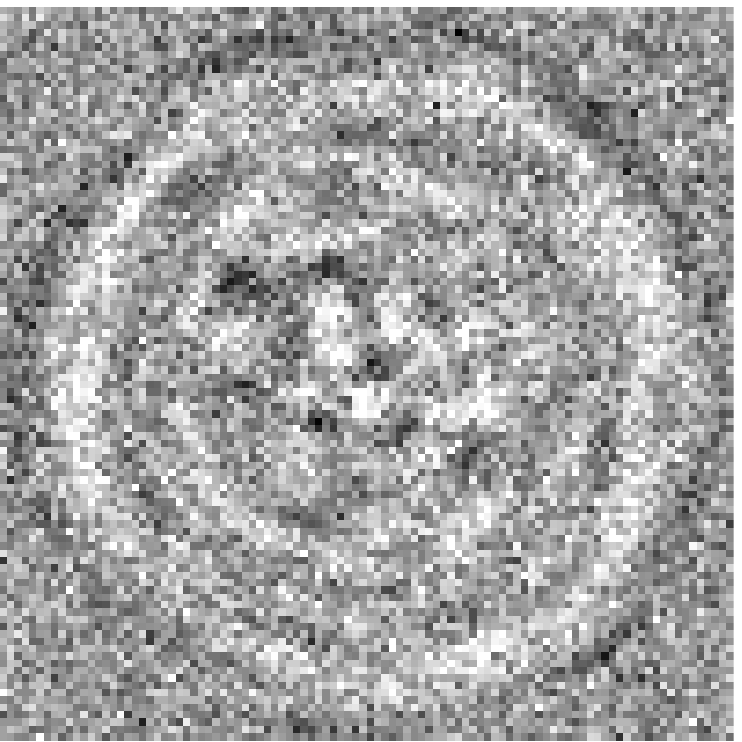}
&
\includegraphics[height=1.1in]{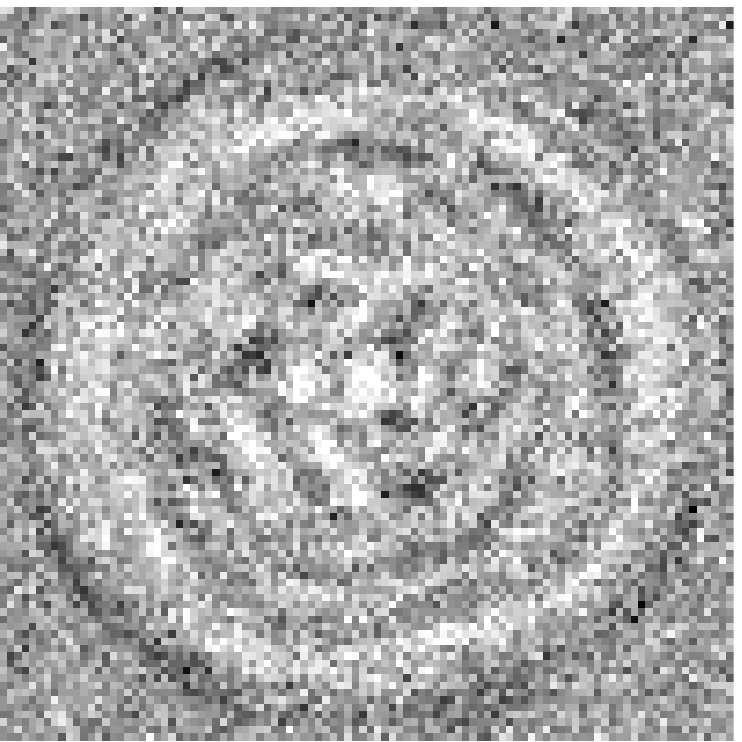}
\end{tabular}
\vspace*{-.2in}
\caption{
\label{fig:simImages}
Three simulated real-space 2-D images of HK97 at SNR 0.25.
The color map saturates for the highest 2\% of the pixel values.
}
\end{figure}

Simulated 2-D projection images, directly in reciprocal space, were generated
from (\ref{eq:2Dmodel}).
There are multiple sources of information from which $\bar\vc$ could be
computed, including from atomic resolution coordinates from x-ray
crystallography experiments.
However, there are few sources of information from which $\mathbf{V}$ could
be computed.
The values of $\bar\vc$ and $\mathbf{V}$ used in the simulation (``ground
truth'') come from the {\heterosymstatistics} reconstruction results from
experimental images for HK97 (Section~\ref{sec:results:HK97:experimental}).
Each of the 1200 images measures $100\times 100$ pixels with a pixel
sampling interval of 5.52~{\AA}.
The signal to noise ratio (SNR) is the ratio of the square of the Euclidean
norm of the $100\times 100$ pixel noise-free image to the variance of the
additive zero-mean Gaussian noise and has value 0.25.
Examples of the simulated real-space 2-D images are shown in
Figure~\ref{fig:simImages}.
\par
\begin{figure}
\begin{tabular}{c@{\hspace{0.05in}}c}
\includegraphics[width=1.65in]{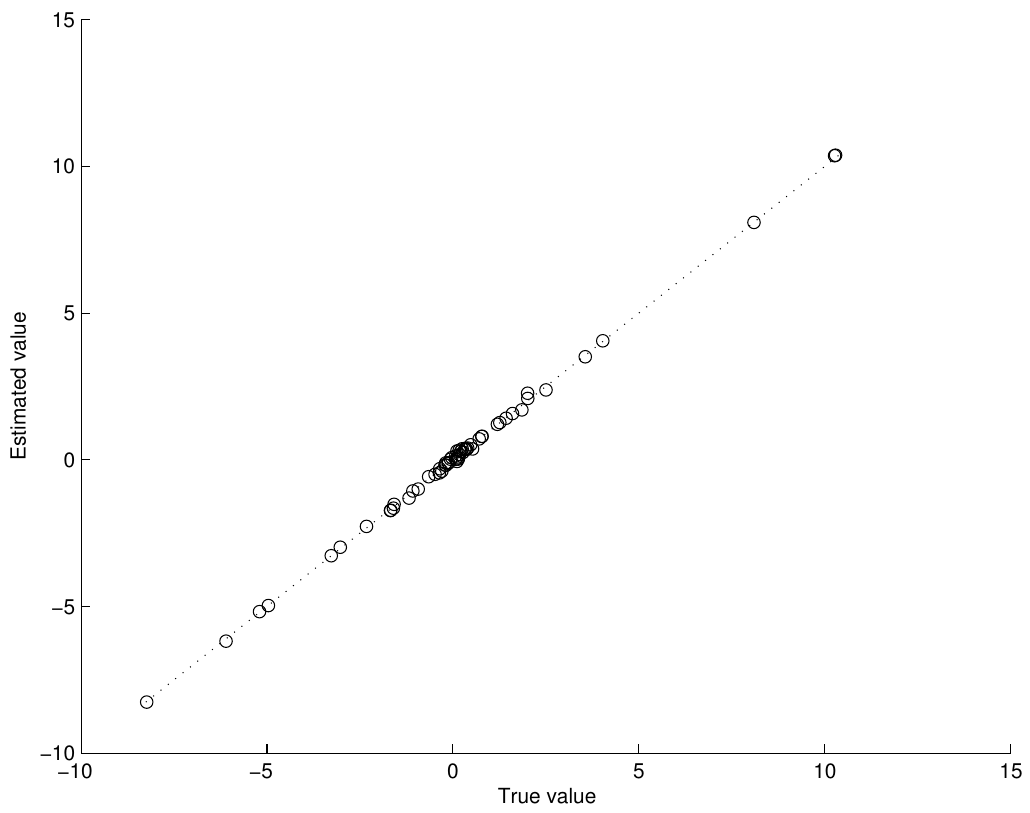}
&
\includegraphics[width=1.65in]{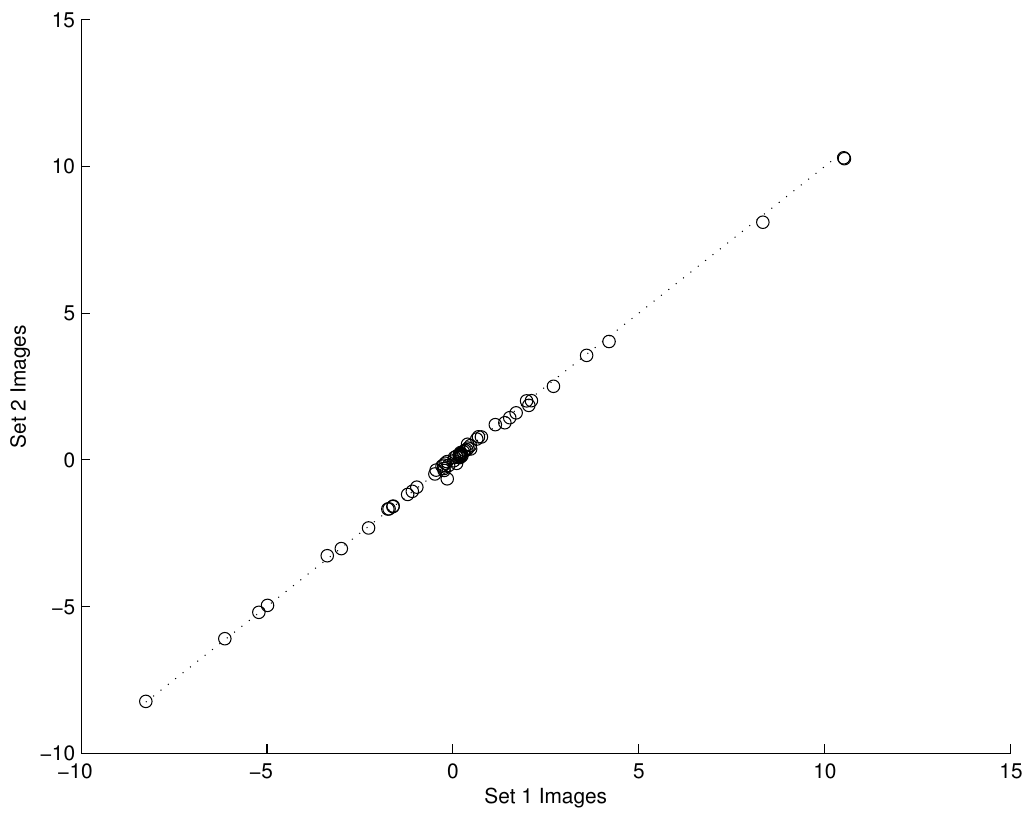}
\\
(a) & (b)
\end{tabular}
\vspace*{-.15in}
\caption{
\label{fig:HK97:synthetic:scatterplot}
\label{fig:HK97:experimentalSymStat:scatterplot}
Scatter plots related to estimates of $\bar{\vc}$.
(a)~True and estimated values of $\bar{\vc}$ for the synthetic-image
calculation of Section~\ref{sec:results:simulated}.
(b)~Estimated values of the $\bar{\vc}$ coefficients
for two separate {\heterosymstatistics} calculations starting from separate
sets of experimental images (Section~\ref{sec:results:HK97:experimental}).
}
\end{figure}

A reconstruction was computed using {\heterosymstatistics}.
Concerning the performance in estimating the mean of the coefficients
($\bar{\vc}$), there are only 60 basis functions with $p=1$ and therefore
only 60 coefficients with nonzero mean.
A scatter plot of the nonzero coefficients is shown in
Figure~\ref{fig:HK97:synthetic:scatterplot}(a).
In addition, $\|\hat{\vc}-\vc\|_1/\|\vc\|_1=0.036$ or 3.6\%.
Concerning the performance in estimating the covariance of the coefficients
($\mathbf{V}$), there are a total of 2020 basis functions and the number of
elements in $\mathbf{V}$ that can be nonzero is too large to show on a
scatter plot.
Therefore, we report $l_1$ norm results:
$\|\mathbf{V}-\hat{\mathbf{V}}\|_1/\|\mathbf{V}\|_1=0.12$ or 12\%.
Resolution is not limited by the amount or the quality of images but rather
by the number of Fourier series terms, i.e., number of terms in the
summation in (\ref{eq:rho:general}), that our computer system is capable of
processing.
The energy curve (\ref{eq:Energy:def}) has decayed by a factor of $10^4$ by
$k=0.04$ $\mbox{\AA}^{-1}$ so it is unlikely that resolution is greater
than 25 {\AA}.
\subsection{Experimental images: HK97}
\label{sec:results:HK97:experimental}
Limitations of our computer system (Section~\ref{sec:MLE}) make it
impossible to process the entire set of 5,978 images in
Ref.~\cite{VeeslerKhayatKrishnamurthySnijderHuangHeckAnandJohnsonHK97Structure2014}
so two sets of 1200 images with no images in common between the two sets
and where each image shows one instance of the particle were randomly
selected for independent processing.
Each image measures $200\times 200$ pixels with a pixel sampling interval
of 2.76~{\AA}.
Two algorithms, {\heterosymparticles} using 1060 basis functions and
{\heterosymstatistics} using 2020 basis functions, were applied to the two
sets of images for a total of four reconstructions.
\par
First, consider the {\heterosymstatistics} results.
The same algorithm and parameter values were used in these calculations as
were used in the synthetic image calculations of
Section~\ref{sec:results:simulated}.
The results are
a scatter plot of non-zero $\hat{\vc}_1$ and $\hat{\vc}_2$
(Figure~\ref{fig:HK97:experimentalSymStat:scatterplot}(b)),
$\|\hat{\vc}_1-\hat{\vc}_2\|_1/(0.5(\|\hat{\vc}_1\|_1+\|\hat{\vc}_2\|))=0.047$
or 4.7\%,
and
$\|\hat{\mathbf{V}}_1-\hat{\mathbf{V}}_2\|_1/(0.5(\|\hat{\mathbf{V}}_1\|_1+\|\hat{\mathbf{V}}_2\|_1))=0.10$ or 10\%.
Resolution is not limited by the amount or the quality of images but rather
by the number of coefficients that our computer system is capable of
processing.
The energy curve (\ref{eq:Energy:def}) has decayed by a factor of $10^4$ by
$k=0.04$ $\mbox{\AA}^{-1}$ so it is unlikely that resolution is greater
than 25 {\AA}.
\par
\begin{figure}
\begin{center}
\includegraphics[width=3.0in]{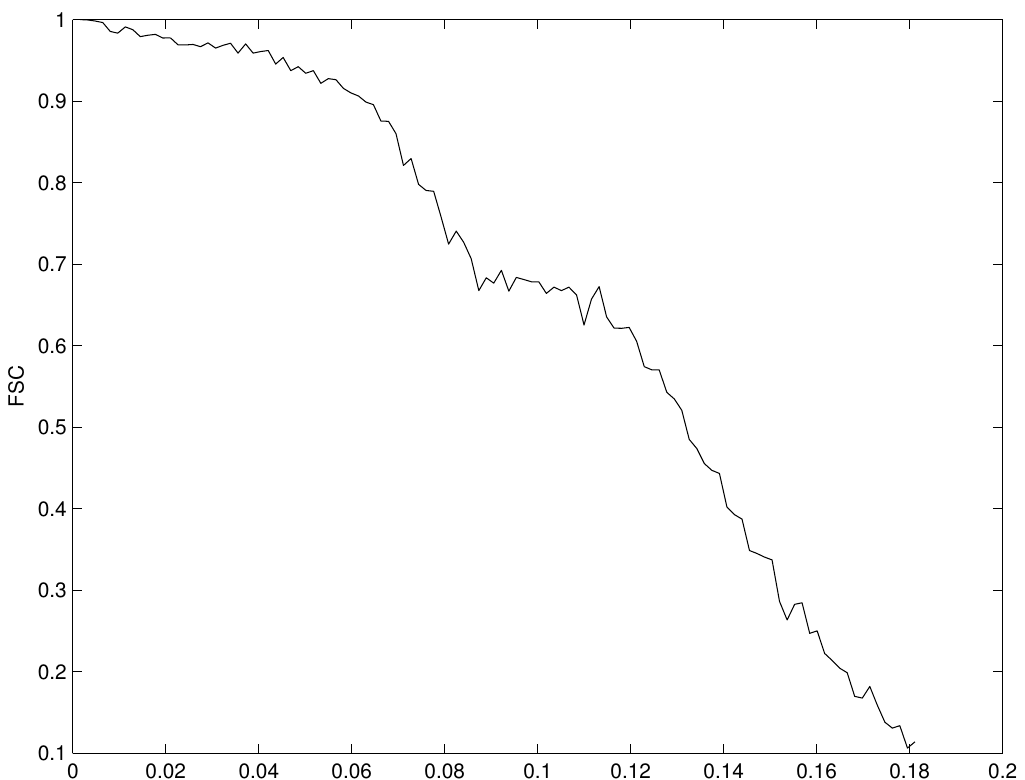}
\\
$k$ ($\mbox{\AA}^{-1}$)
\end{center}
\vspace*{-.2in}
\caption{
\label{fig:HK97:experimentalSymPart:FSC}
FSC curve (command {\tt calcfsc} of
EMAN2~\cite{TangPengBaldwinMannJiangReesLudtkeJSB2007EMAN2})
comparing two {\heterosymparticles} calculations based on
separate sets of experimental images using initial conditions derived from
the separate sets of images.
The resolution (the inverse of the value of $k$ where the curve crosses
0.5) is $1/0.131=7.63$ {\AA}.
}
\end{figure}

Second, consider the {\heterosymparticles} results.
An FSC curve (command \url{calcfsc} of
EMAN2~\cite{TangPengBaldwinMannJiangReesLudtkeJSB2007EMAN2})
for two estimates of the mean (estimates of $\bar\rho(\vx)$) computed from
separate sets of experimental images using initial conditions derived from
the separate sets of images is shown in
Figure~\ref{fig:HK97:experimentalSymPart:FSC}.
The resolution, which is the inverse of the value of $k$ where the curve
crosses 0.5, is $1/0.131=7.63$ {\AA}.
\par
\begin{figure}
	\begin{center}
		\begin{tabular}{cc}
			\includegraphics[width=3.8cm,trim={.8in .45in .3in 1.5in},clip]{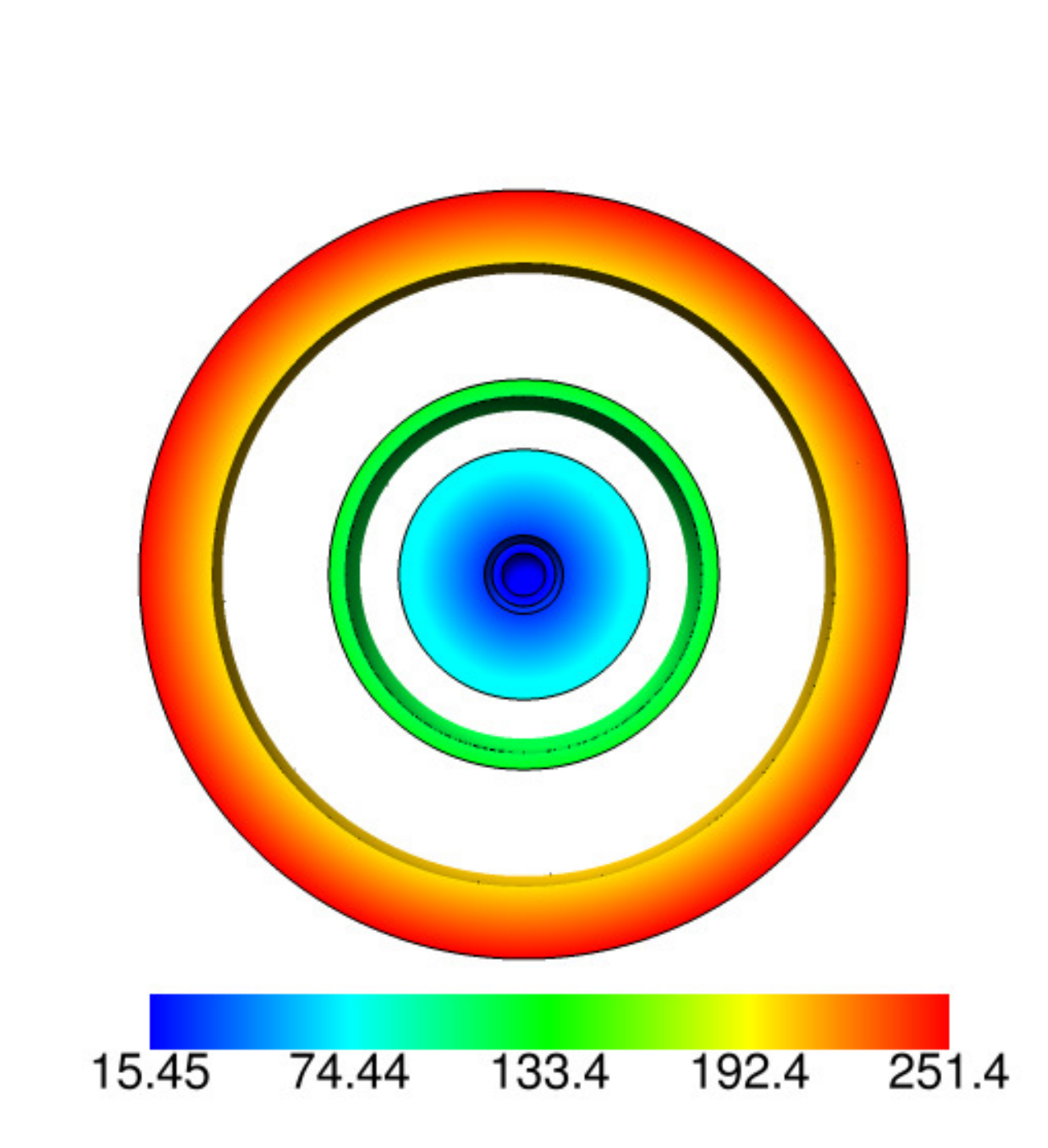}
			&
			\includegraphics[width=4.0cm,trim={.8in .45in .3in 1.5in},clip]{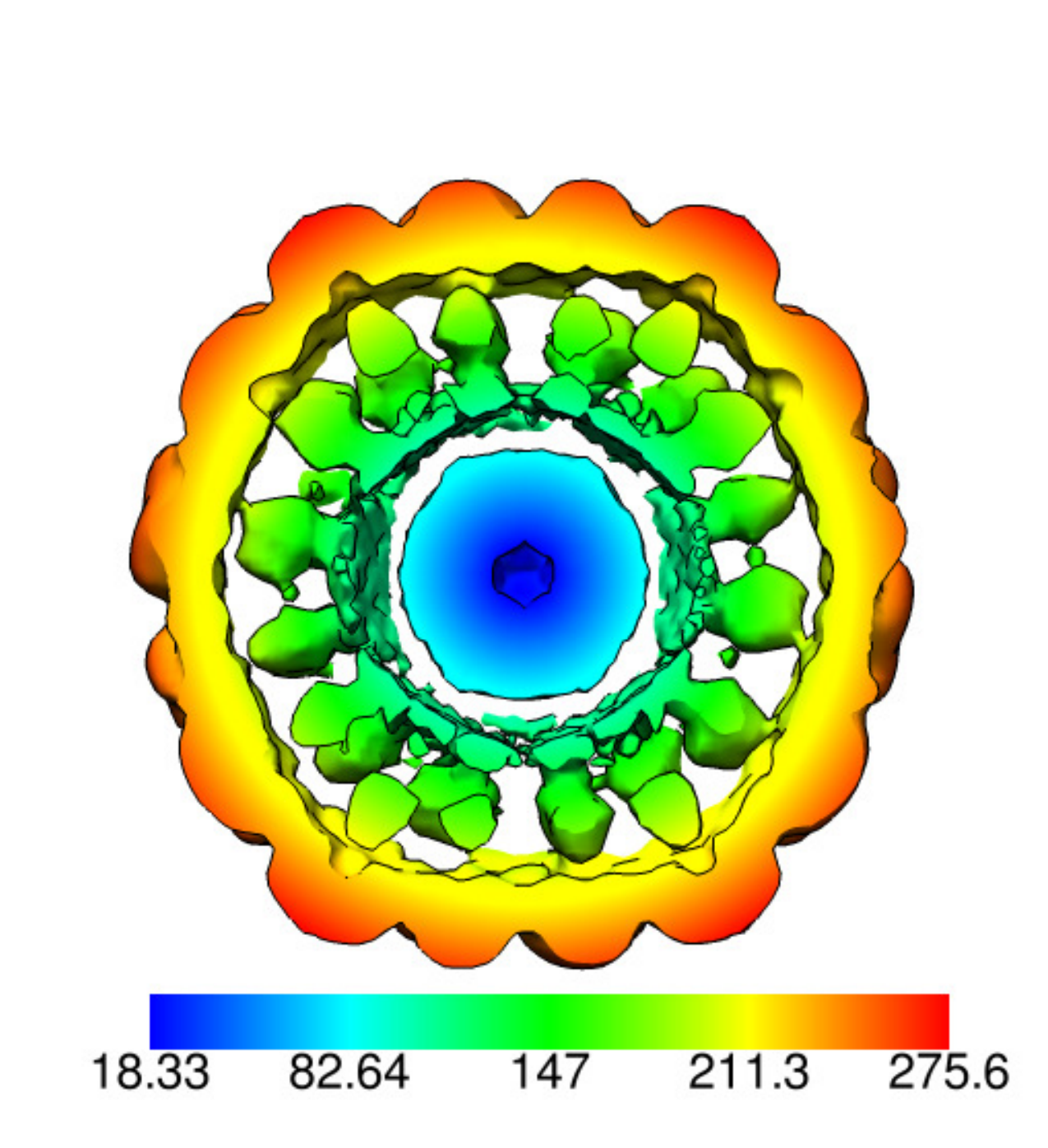}
		\end{tabular}
	\end{center}
	\vspace*{-.1in}
	\caption{
		\label{fig:HK97:homogeneousreconstructions}
		Visualization of {\withpro} $\bar\rho(\vx)$ estimates from {\heterohomo}:
		cross sections of $\bar\rho(\vx)$ estimate (contour level $5\times
		10^{-4}$) colored by $\bar\rho(\vx)$ estimate.
		Left panel: $l=0$.
		Right panel: $l=55$.
		Both panels: $p=1$, $q\in\{1,\dots,20\}$, and $R_2=280~\mbox{\AA}$.
		Visualization by UCSF
		Chimera~\cite{PettersenHuangCouchGreenblattMengFerrin2004}.
	}
\end{figure}
\begin{figure}
	\begin{center}
		\begin{tabular}{cc}
			\includegraphics[width=3.8cm,trim={.95in .45in 0 1.5in},clip]{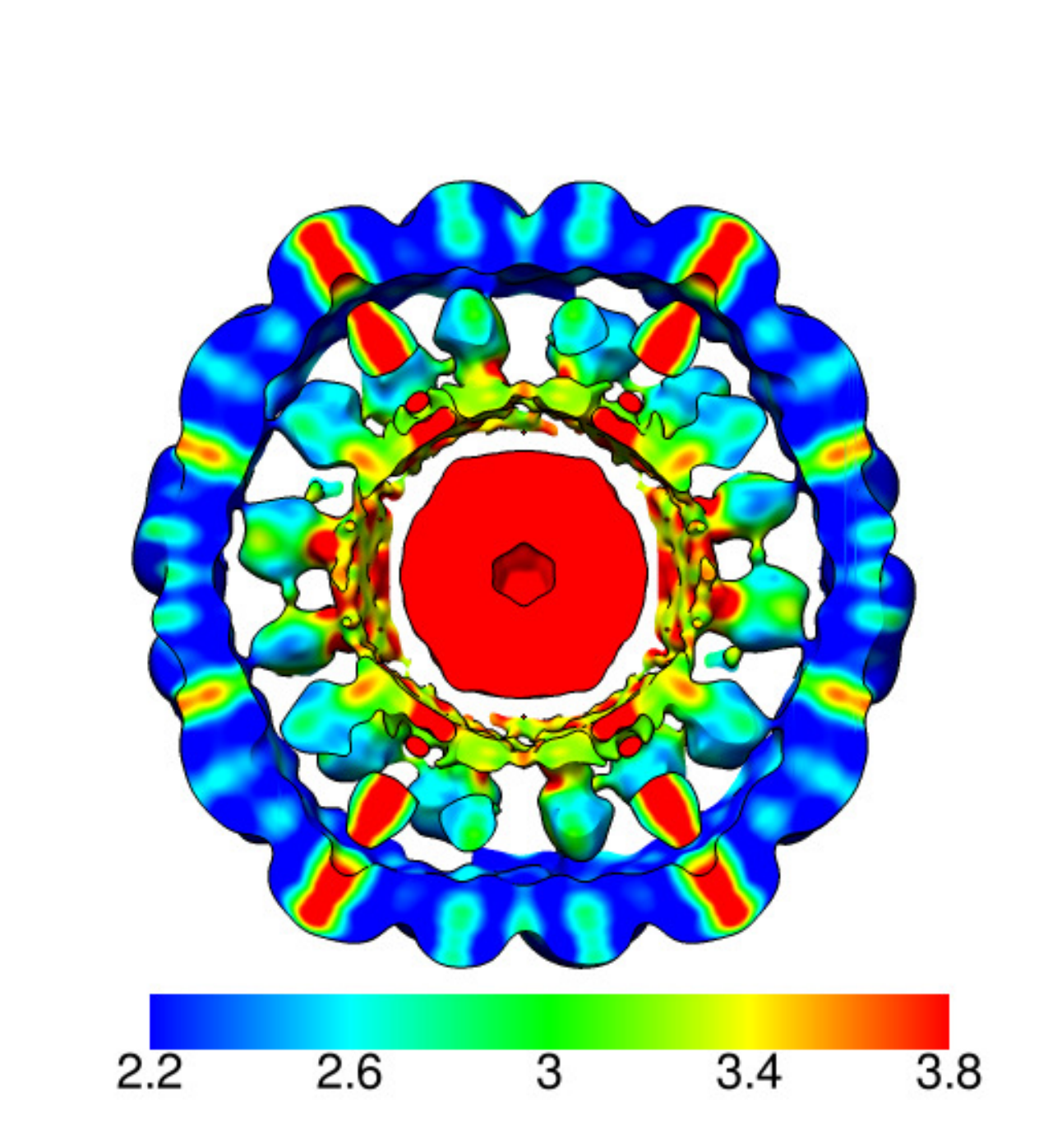}
			&
			\includegraphics[width=3.8cm,trim={.95in .45in 0 1.5in},clip]{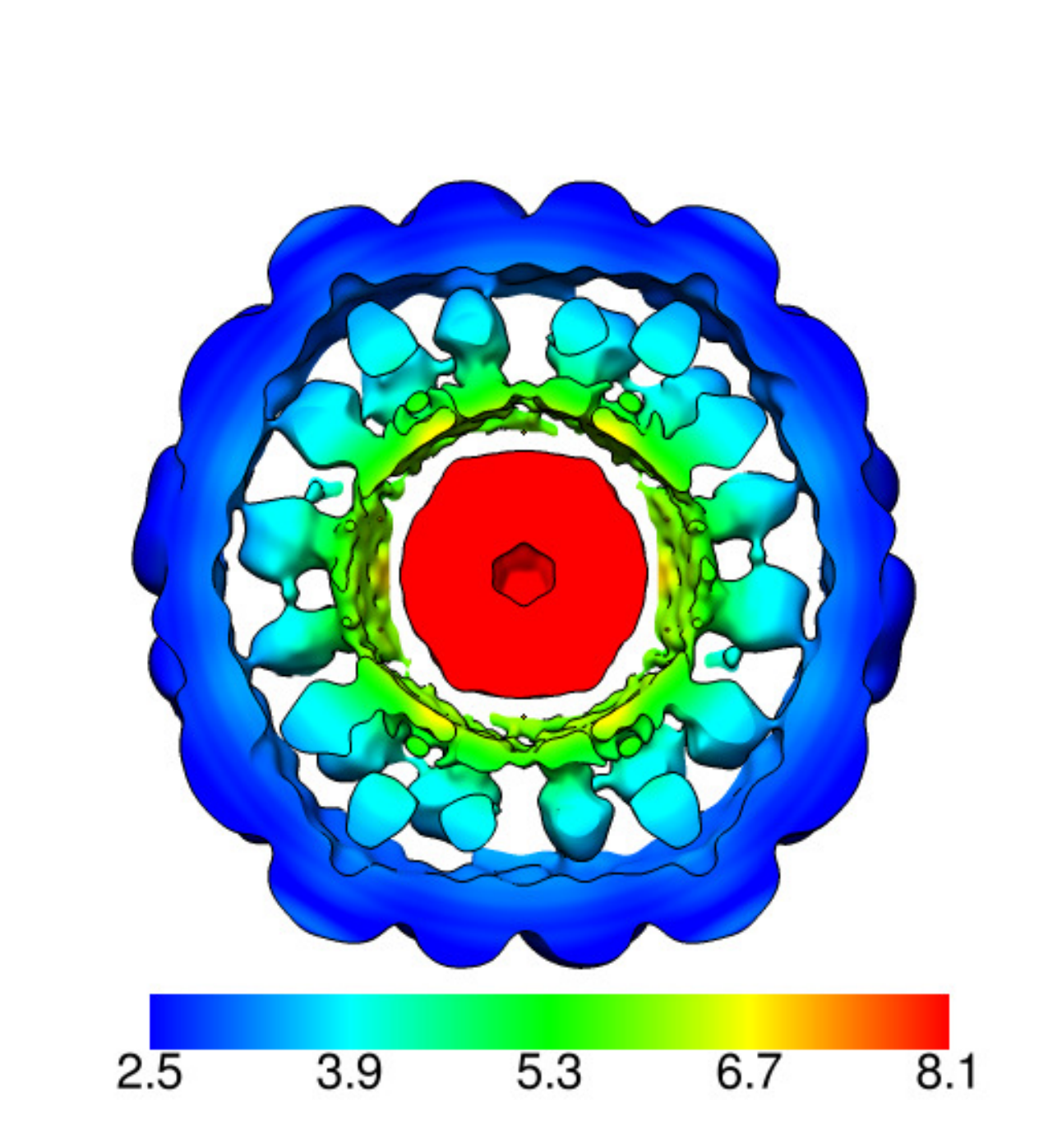}
			\\
			(a) symmetric particles
			&
			(b) symmetric statistics
			\\
			(\heterosymparticles)
			&
			(\heterosymstatistics)
		\end{tabular}
	\end{center}
	\vspace*{-.1in}
	\caption{
		\label{fig:rhobar_d}
		Joint visualization of $\bar\rho(\vx)$ and $s(\vx)$ estimates: cross
		sections of {\withpro} $\bar\rho(\vx)$ estimate (contour level $5\times
		10^{-4}$) from {\heterosymparticles} colored by $s(\vx)$ estimate from
		{\heterosymparticles} (Panel~(a)) or {\heterosymstatistics} (Panel~(b)).
		Panel~(a): $p=1$ and $l\in\{0,1,\dots,55\}$.
		Panel~(b): $p\in\{1,\dots,\Nirrep=5\}$ and $l\in\{0,1,\dots,10\}$.
		Both panels: $q\in\{1,2,\dots,20\}$ and $R_2=280~\mbox{\AA}$.
		All markings are scaled by $10^{-3}$.
		Visualization by UCSF
		Chimera~\cite{PettersenHuangCouchGreenblattMengFerrin2004}.
	}
\end{figure}

Third, compare the {\heterosymparticles} and {\heterosymstatistics}
reconstructions.
In particular, in Figure~\ref{fig:HK97:homogeneousreconstructions}, the
mean $\bar\rho(\vx)$ is visualized for the {\heterohomo} reconstructions
that are Steps~(i) and (ii) of the three-step reconstruction process.
Furthermore, in Figure~\ref{fig:rhobar_d}, the mean $\bar\rho(\vx)$ and
the standard deviation $s(\vx)$ estimates are jointly visualized for both
both heterogeneous reconstructions by
using the $\bar\rho(\vx)$ estimate from {\heterosymparticles} to define a
shape and the $s(\vx)$ estimate from either {\heterosymparticles} or
{\heterosymstatistics} to color the shape. As is shown in Figure~\ref{fig:rhobar_d}(a) (the red radial streaks),
{\heterosymparticles} estimates a standard deviation function that is
organized in radially-organized regions of high standard deviation that
lie along all three types of symmetry axis, 2-, 3-, and 5-fold symmetry
axes.
The highest values are along the 5-fold axes, the lowest along the 2-fold
axes, with intermediate values along the 3-fold axes.
It is difficult to understand this result in terms of the biology of the
particle because the particle is organized in annular shells.
In particular, from outside to inside, the {\withpro} particle is composed
of a shell of well-ordered capsid protein, poorly-ordered capsid
$\delta$-domain and protease protein, and poorly ordered DNA genome.
On the other hand, {\heterosymstatistics} estimates a standard deviation
function that is organized in annular regions that matches the physical
organization of the particle.
Using a different approach, specifically, a resampling approach, but still
assuming that each instance of the particle is symmetric, this same type of
behavior, maxima in the neighborhood of symmetry axes, is well-known to be
a common anomalous result~\cite[p.~173]{Ludtke.MethodsEnzymology.2016} from
standard software systems, e.g.,
EMAN2~\cite{TangPengBaldwinMannJiangReesLudtkeJSB2007EMAN2}.
Our results with {\heterosymstatistics} suggest a possible solution to this
problem.
Specifically, if each reconstruction from the resampled data was computed
with no symmetry constraints, then the sample mean and covariance of the
reconstructions might have no anomalous maxima and (at least approximately)
show the symmetry.
However, this solution would require a large increase in both computation and
data reflecting the difference between the number of variance parameters for
a {\heteroasymmetry} calculation versus the number of variance parameters
for a {\heterosymstatistics} calculation in
Figure~\ref{fig:computationalcomplexity}.
\begin{figure*}
	\begin{center}
		\begin{tabular}{c@{\hspace{0.05in}}c@{\hspace{0.05in}}c@{\hspace{0.05in}}c}
			\includegraphics[width=1.5in,trim={7.0in 1.2in 9.2in 1.7in},clip]{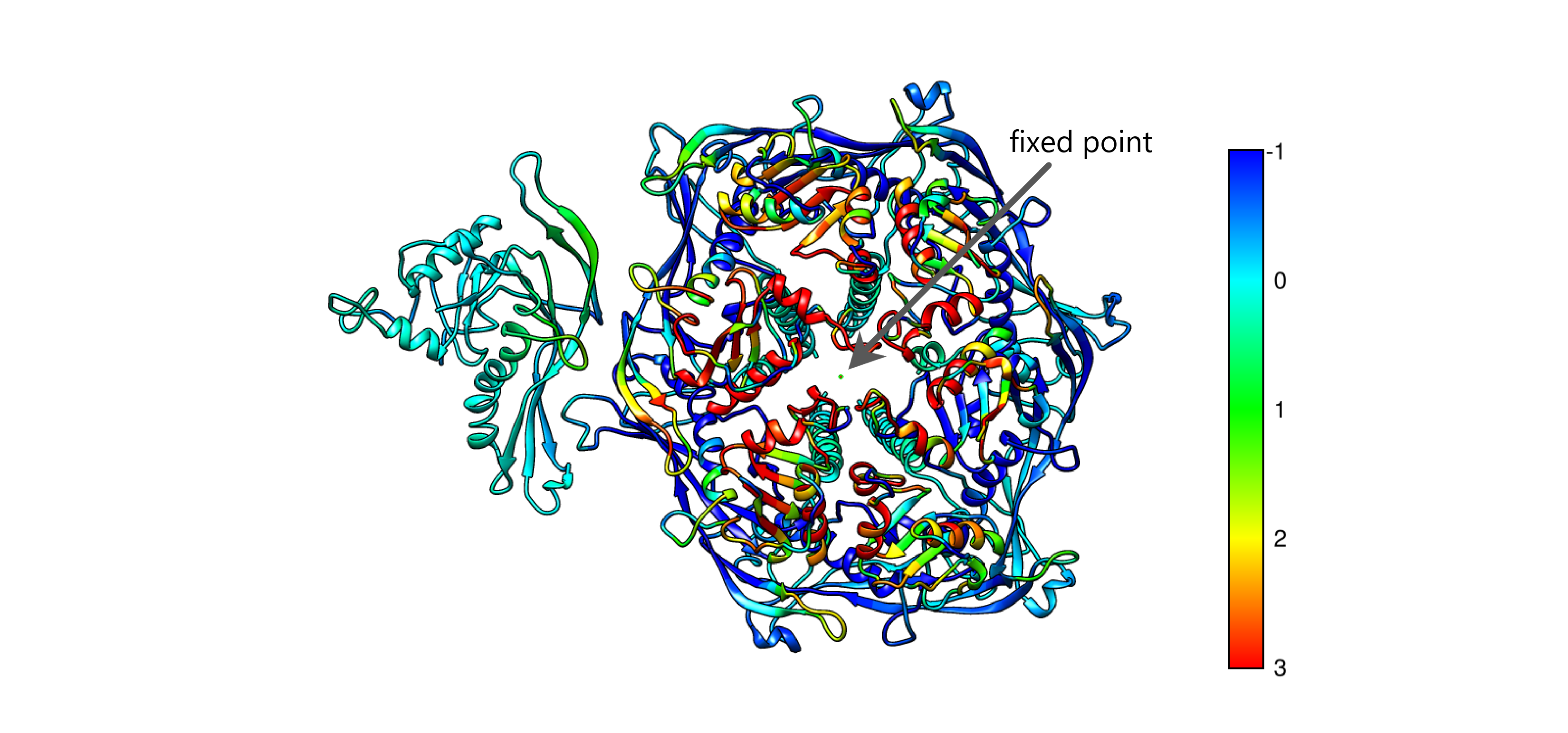}
			&
			\includegraphics[width=1.5in,trim={7.0in 1.2in 9.6in 2.4in},clip]{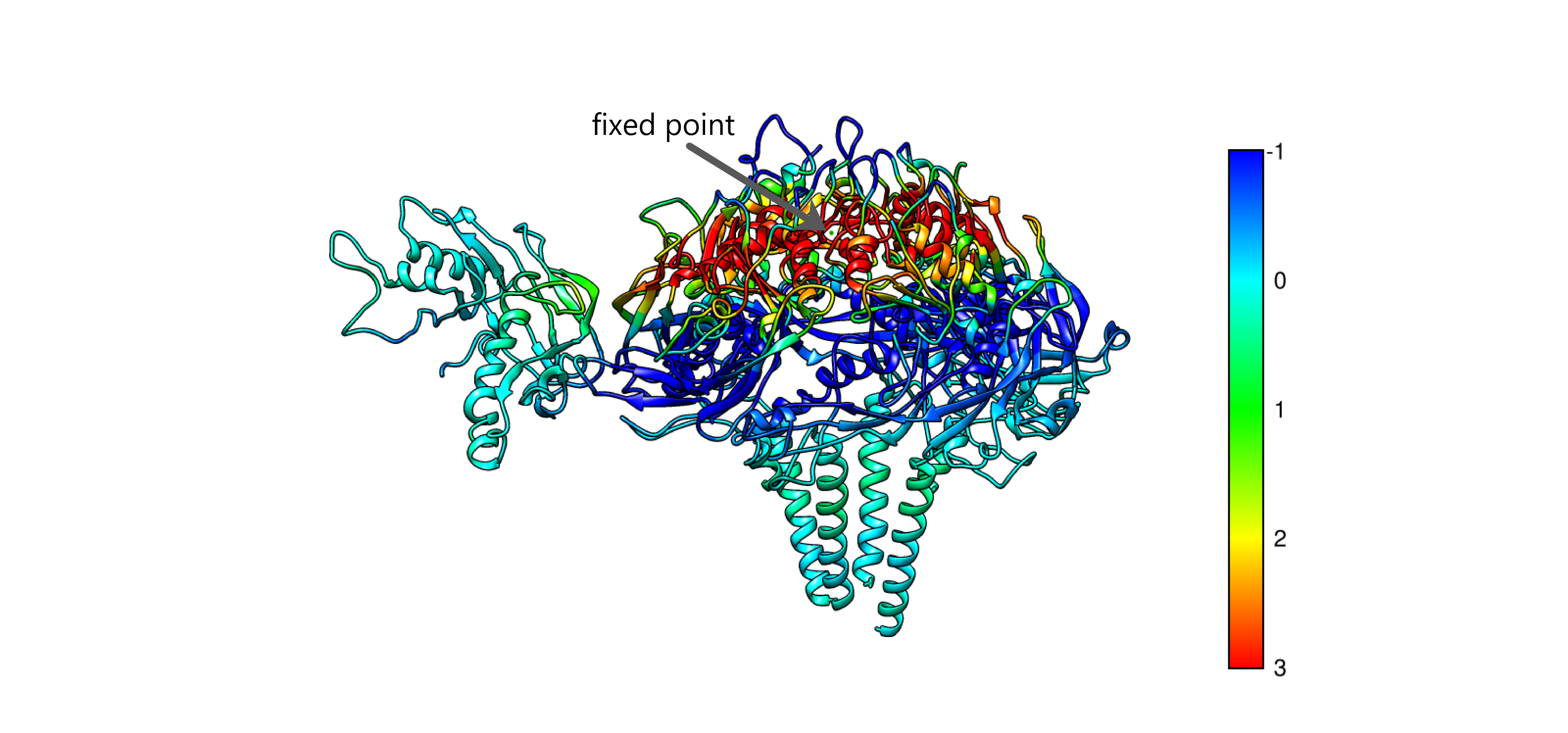}
			&
			\includegraphics[width=1.5in,trim={7.0in 1.4in 8.9in 1.7in},clip]{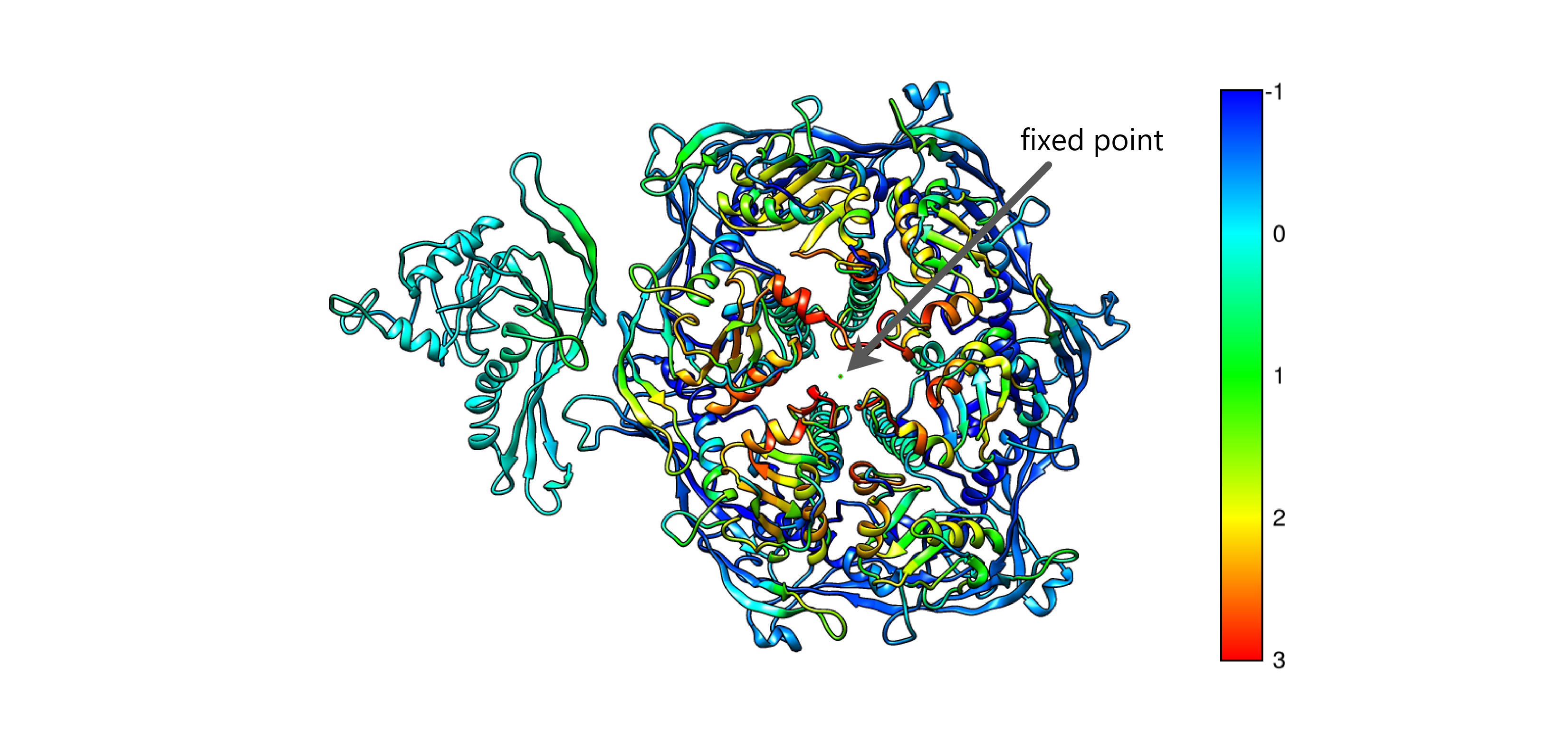}
			&
			\includegraphics[width=1.9in,trim={7.0in 1.4in 6.0in 1.9in},clip]{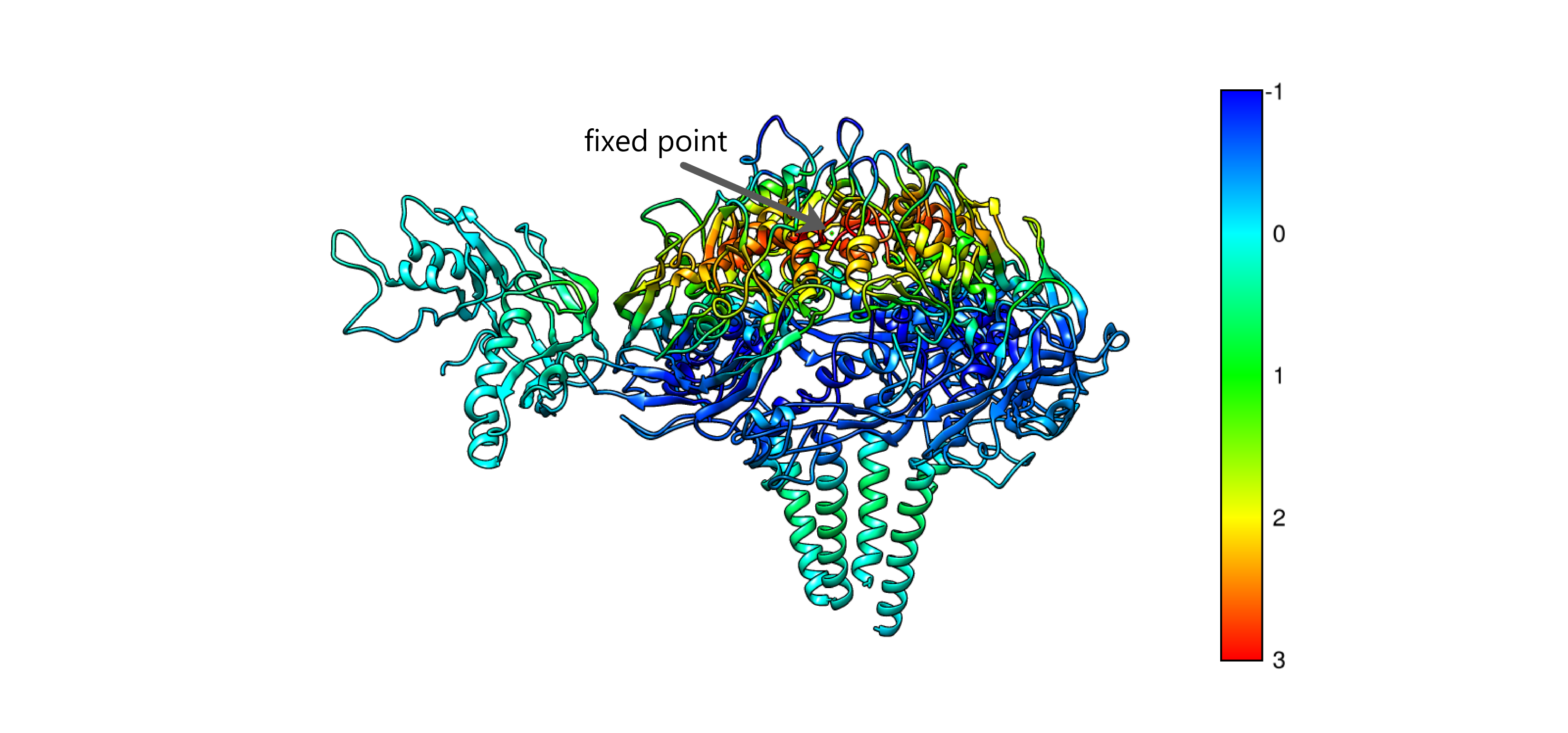}
			\\
			(a) {\withoutpro} & (b) {\withoutpro}
			&
			(c) {\withpro} & (d) {\withpro}
		\end{tabular}
	\end{center}
	\vspace*{-.1in}
	\caption{
		\label{fig:SymStat:allosteric}
		Visualization of $\CrhoREAL(\vx_1,\vx_2)$ from {\heterosymstatistics} via
		ribbon diagrams.
		Panels~(a) and~(b): The ribbon diagram showing seven copies of the capsid
		peptide for {\withoutpro} where the color is the value of
		$\CrhoREAL(\vx_1,\vx_2)$ where $\vx_2$ is fixed at the indicated location
		while $\vx_1\in\mathbb{R}^3$.
		Panels~(c) and~(d): The same as Panels~(a) and~(b) except for {\withpro}
		instead of {\withoutpro}.
		Panels~(a) and~(c): From outside of the particle looking toward the
		particle's center.
		Panels~(b) and~(d): Tangential to the particle's capsid, the external
		(internal) surface is at the top (bottom) of the image.
		The protease binds to the $\delta$ domain which is the extended alpha helices
		at the internal surface in the side view.
		The absence of the protease makes the surface of the capsid rigid.
		Specifically,
		$\CrhoREAL(\vx_1,\vx_2=\vx_{\rm fixed})$ and $\vx_{\rm fixed}$ is on the
		5-fold symmetry axis near the external surface of the particle as is
		indicated by the arrows.
		If the region of the capsid around $\vx_{\rm fixed}$ is rigid then
		$\CrhoREAL$ will be large, i.e., red, in that region.
		This is what is seen in {\withoutpro}, especially the side view of
		Panel~(b).
		On the other hand, if the region of the capsid around $\vx_{\rm fixed}$ is
		less rigid then $\CrhoREAL$ will be smaller, i.e., yellow, green, ..., down
		to light blue ($\CrhoREAL=0$) in that region.
		This is what is seen in {\withpro}, especially the side view of
		Panel~(d), where the region that was mostly red in {\withoutpro} is mostly
		yellow and green in {\withpro}.
		Since the effect is distant from the binding site of the protease, this is
		an example of an allosteric interaction.
		There is no corresponding 6-D visualization of {\heterosymparticles} because
		$\CrhoREAL(\vx_1,\vx_2)$ for {\heterosymparticles} is periodic and for that
		reason is not interpretable.
		The reason that $\CrhoREAL(\vx_1,\vx_2)$ for {\heterosymparticles} is
		periodic is the fact that for {\heterosymparticles}, all of the basis
		functions in (\ref{eq:Crho}) are invariant under all of the symmetries of
		the icosahedral rotational point group.
		Visualization by UCSF
		Chimera~\cite{PettersenHuangCouchGreenblattMengFerrin2004}.
		For {\withpro} ({\withoutpro}), 95.91\% (73.91\%) of the amino acid
		residues have covariance values that are between the upper and lower bounds
		of the colorbar.
		All markings are scaled by $10^{-6}$.
	}
\end{figure*}
\par
Finally, in Figure~\ref{fig:SymStat:allosteric}, the covariance function 
of the electron scattering intensity, $\CrhoREAL(\vx_1,\vx_2)$ is
visualized for {\heterosymstatistics} reconstructions of two particles,
{\withpro} and a second particle that entirely lacks the protease which is
denoted by {\withoutpro}.
{\withpro} and {\withoutpro} both trap at the same step in maturation
because neither can cleave the capsid protein.
To the best of our knowledge, this is the first publication which presents
a full 6-D covariance function in the presence of symmetry and the
visualization and interpretation of 6-D of information is challenging.
In the visualization, one of the locations ($\vx_2$) is fixed at the
indicated position and the other location ($\vx_1$) is allowed to vary in
$\mathbb{R}^3$.
The resulting 3-D cube is shown as the color of a ribbon diagram showing
each of the seven copies of the capsid peptide.
The absence of the protease (which is bound in an unknown location on the
inner surface of the capsid) makes the surface of the capsid rigid (large
covariance values as indicated by the band of red).
Since the effect is distant from the binding site of the protease, this is
an example of an allosteric interaction.
\subsection{Experimental images: N$\omega$V}
\label{sec:results:NwV:experimental}
\begin{figure}
	\begin{center}
		\begin{tabular}{cc}
			\includegraphics[width=3.8cm,trim={1.3in 1.0in .7in 2.5in},clip]{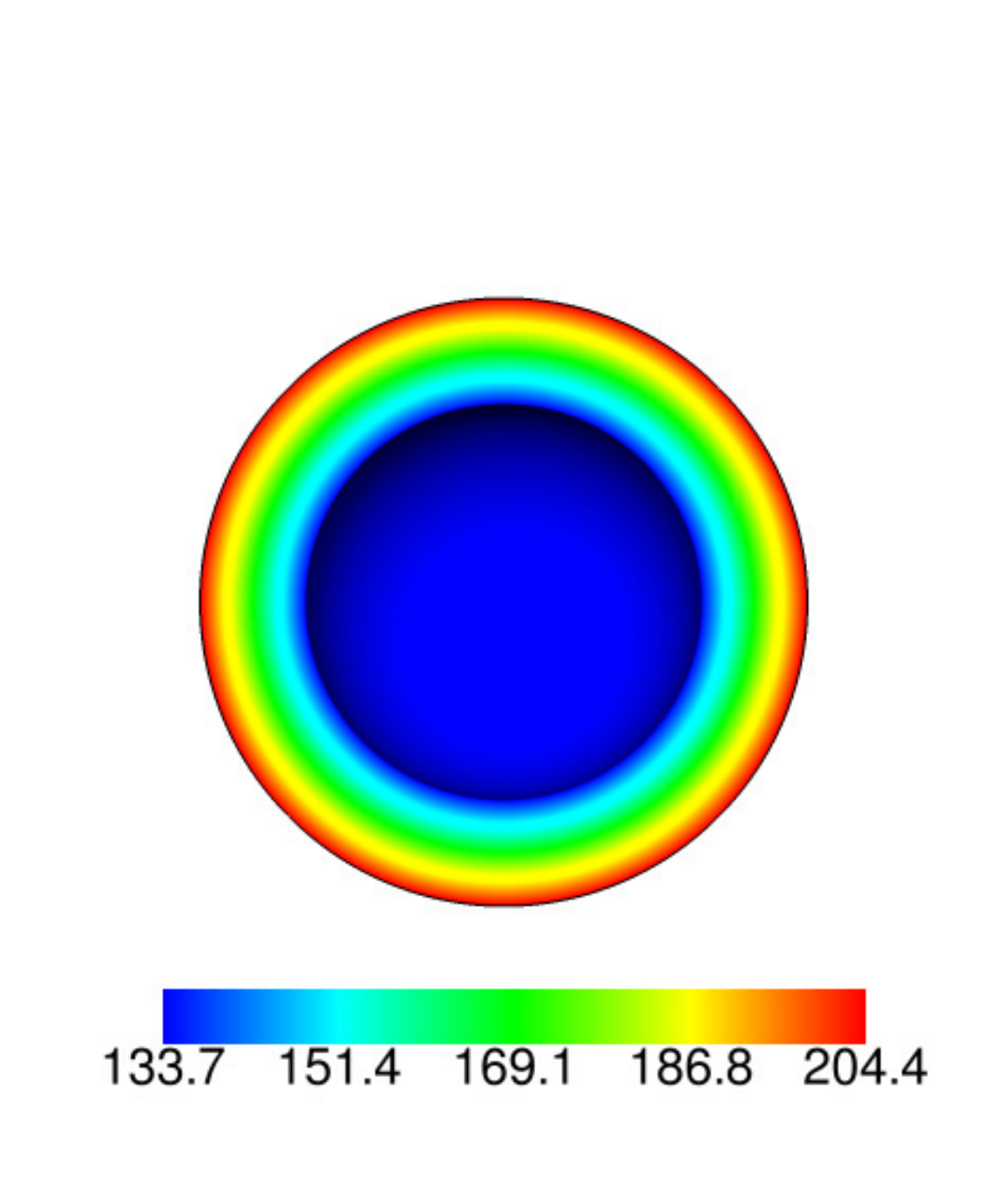}
			&
			\includegraphics[width=3.8cm,trim={1.3in 1.0in .7in 2.5in},clip]{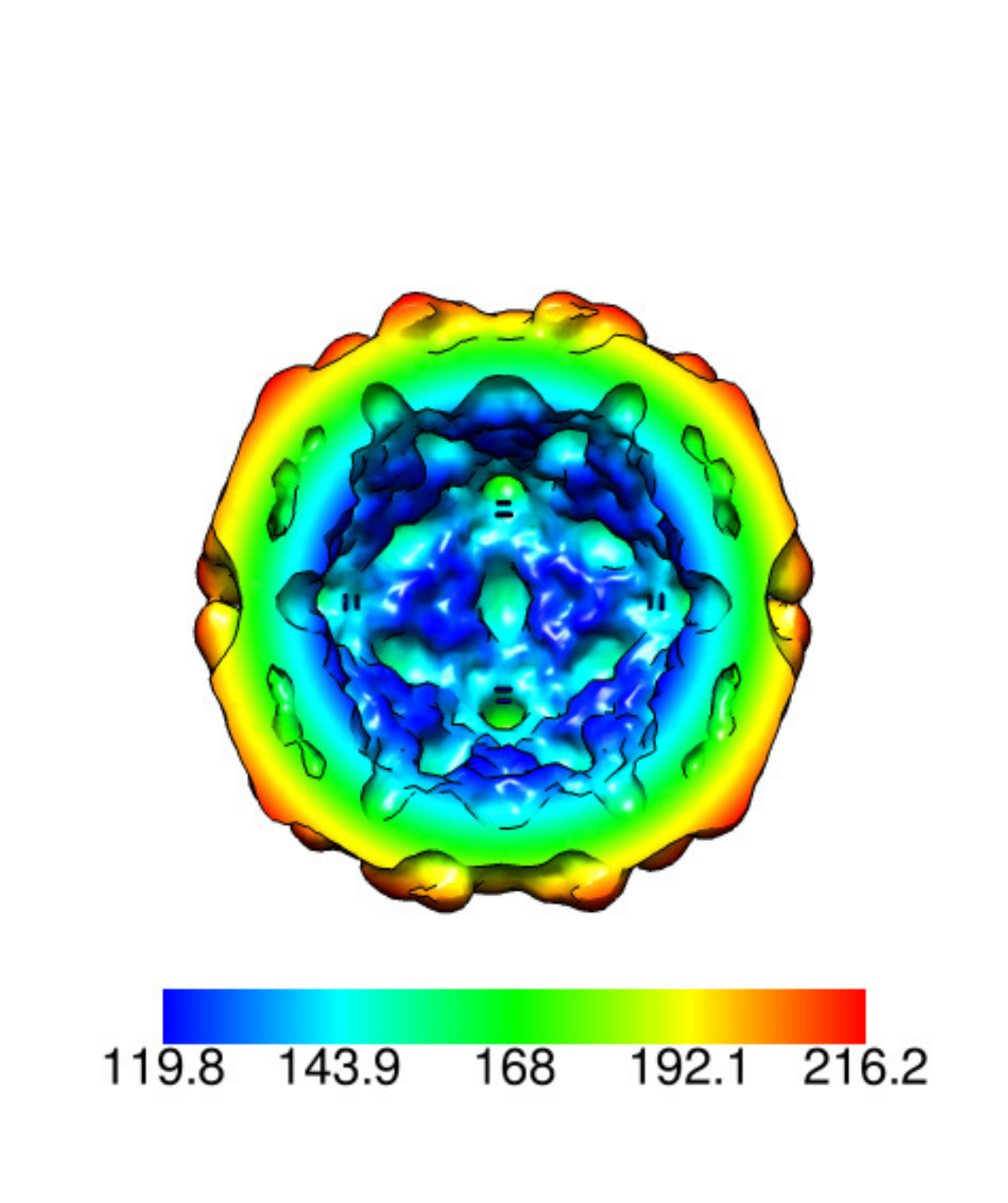}
		\end{tabular}
	\end{center}
	\vspace*{-.1in}
	\caption{
		\label{fig:NwV:homogeneousreconstructions}
		Visualization of N$\omega$V $\bar\rho(\vx)$ estimates from {\heterohomo}:
		cross sections of $\bar\rho(\vx)$ estimate (contour level $5\times
		10^{-4}$) colored by $\bar\rho(\vx)$ estimate.
		Left panel: $l=0$.
		Right panel: $l=46$.
		Both panels: $p=1$, $q\in\{1,\dots,17\}$, and $R_2=230~\mbox{\AA}$.
		This figure is the N$\omega$V analog of
		Figure~\ref{fig:HK97:homogeneousreconstructions}.
		Visualization by UCSF
		Chimera~\cite{PettersenHuangCouchGreenblattMengFerrin2004}.
	}
\end{figure}

\begin{figure}
	\begin{center}
		\begin{tabular}{cc}
			\includegraphics[width=3.8cm,trim={1.3in 1.0in .9in 2.5in},clip]{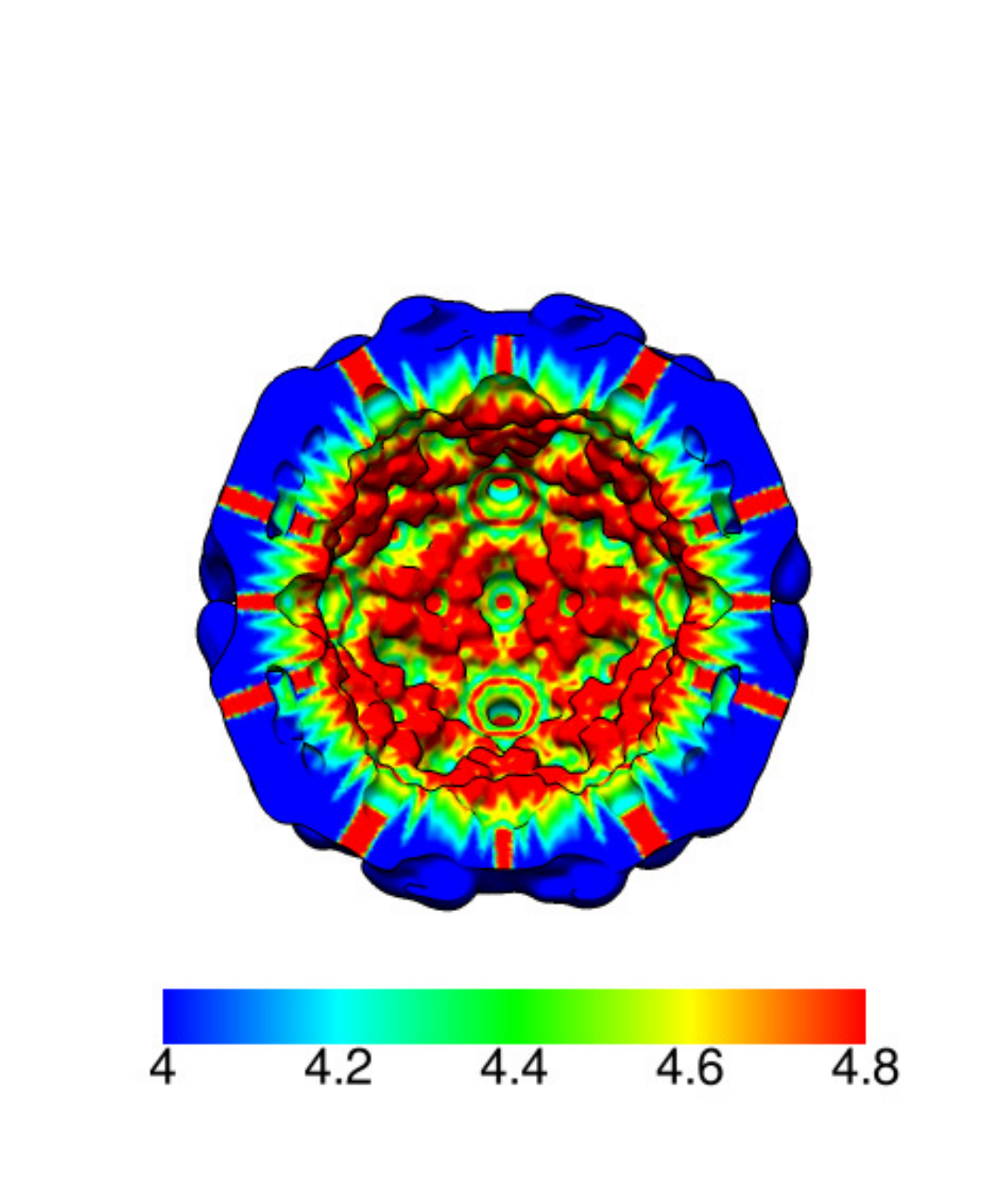}
			&
			\includegraphics[width=3.8cm,trim={1.3in 1.0in .9in 2.5in},clip]{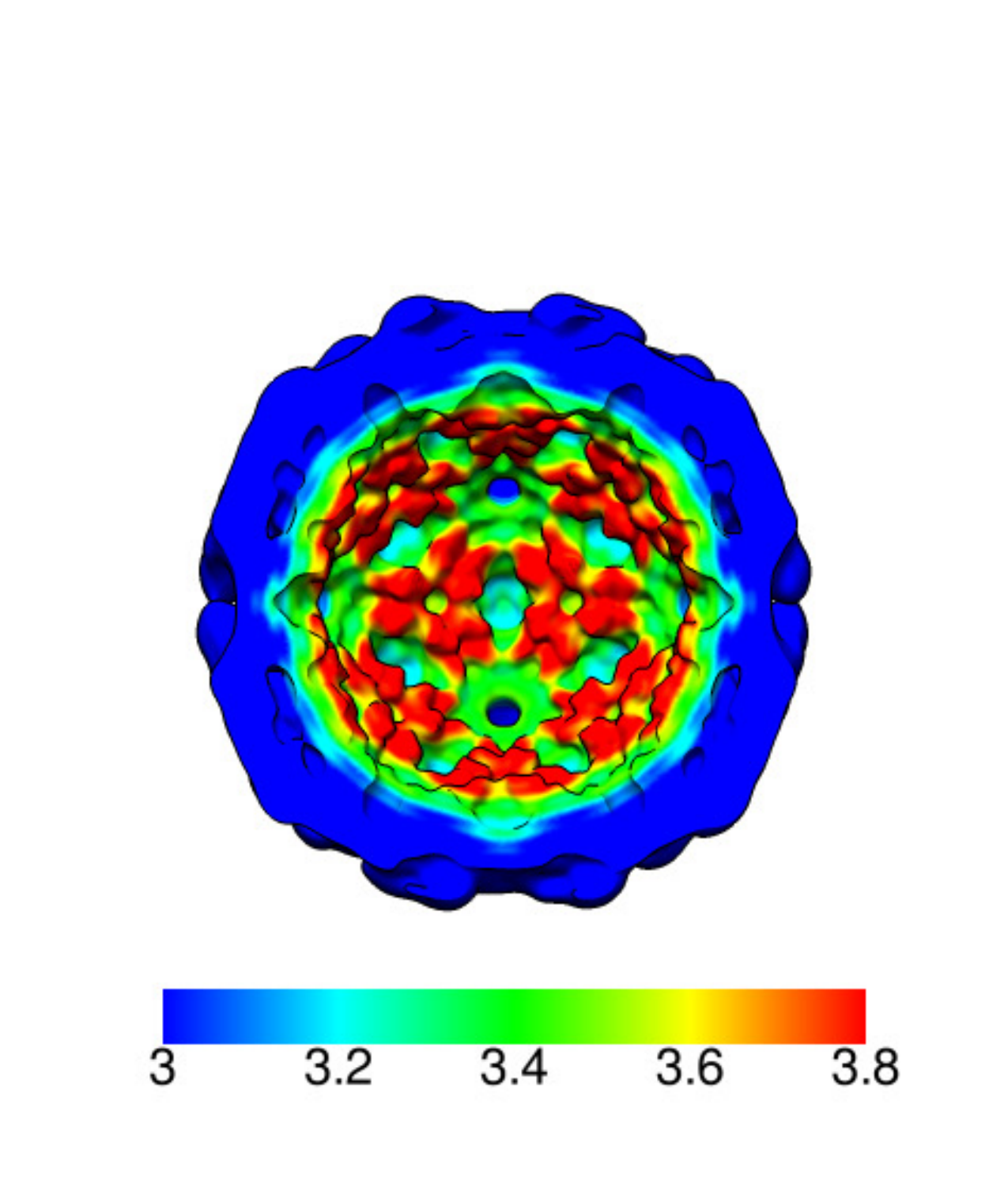}
			\\
			(a) symmetric particles
			&
			(b) symmetric statistics
			\\
			(\heterosymparticles)
			&
			(\heterosymstatistics)
		\end{tabular}
	\end{center}
	\vspace*{-.1in}
	\caption{
		\label{fig:NwV}
		Joint visualization of $\bar\rho(\vx)$ and $s(\vx)$ estimates: cross
		sections of N$\omega$V $\bar\rho(\vx)$ estimate (contour level $5\times
		10^{-4}$) from {\heterosymparticles} colored by $s(\vx)$ estimate from
		{\heterosymparticles} (Panel~(a)) or {\heterosymstatistics} (Panel~(b)).
		Panel~(a): $p=1$ and $l\in\{0,1,\dots,46\}$.
		Panel~(b): $p\in\{1,\dots,\Nirrep=5\}$ and $l\in\{0,1,\dots,10\}$.
		Both panels: $q\in\{1,2,\dots,17\}$ and $R_2=230~\mbox{\AA}$.
		All markings are scaled by $10^{-3}$.
		This figure is the N$\omega$V analog of
		Figure~\ref{fig:rhobar_d}.
		Visualization by UCSF
		Chimera~\cite{PettersenHuangCouchGreenblattMengFerrin2004}.
		In order to better visualize the inner surface of the capsid, a ball of
		radius 120~{\AA} containing the genome has been removed.
		The genome structure varies greatly from particle to particle leading to
		the high standard deviation values seen on the inner surface of the capsid.
	}
\end{figure}

The calculations are similar to those of
Section~\ref{sec:results:HK97:experimental} on HK97.
Due to limitations of our computer system (Section~\ref{sec:MLE}), we again
process only 1200 images.
The {\heterohomo} followed by {\heterosymparticles} reconstruction is
unchanged from Section~\ref{sec:results:HK97:experimental}.
The {\heterohomo} followed by {\heterosymstatistics} demonstrate the
ability of the ideas, algorithms, and software to compute reconstructions
in a spherical annulus ($50~\mbox{\AA}\le \|\vx\| \le 230~\mbox{\AA}$)
rather than simply a ball.
The range of $l$ and $q$ have been adjusted, as is described in the
captions of Figures~\ref{fig:NwV:homogeneousreconstructions}
and~\ref{fig:NwV}, so that the resolution is the same as in the larger HK97
reconstruction.
\par
We focus on whether symmetric statistics is a better mathematical model
than the more commonly used symmetric particles.
The qualitative aspects of Figure~\ref{fig:NwV} for the N$\omega$V
calculations closely match those of Figure~\ref{fig:rhobar_d} for the HK97
calculations.
In both cases, it is difficult to explain the high standard deviation
values that occur along the symmetry axes in the {\heterosymparticles}
(Panel~(a)) calculations.
On the other hand, in both cases the radial organization of the standard
deviation values in the {\heterosymstatistics} (Panel~(b)) matches the
known arrangement of the particles' constituents.
As was discussed relative to the HK97 calculations, the high standard
deviation anomaly is well
known~\cite[p.~173]{Ludtke.MethodsEnzymology.2016} and these results
suggest a possible solution.
\section{Conclusion}
\label{sec:conc}
In this paper we formulate and solve 3-D modeling and reconstruction
problems for stochastic signals where the statistics of the stochastic
signal are required to have a symmetry.
The work is motivated by cryo-electron microscopy, which is of great
importance in structural biology and of growing importance in the materials
science of nanoscale particles.
The focus is on the small-scale heterogeneity that might be characterized
for each class of a multi-class reconstruction problem, although the
examples described actually are one-class problems.
\par
Previous algorithms assumed that the realizations of the stochastic signal
have the symmetry.
While this implies that the statistics have the symmetry, it is a
sufficient condition rather than a necessary condition and it may be the
cause of inaccurate results for the second order statistics of the electron
scattering intensity on and near symmetry
axes~\cite[p.~173]{Ludtke.MethodsEnzymology.2016}.
Since symmetry axes are often locations of important biological events,
e.g., the 5-fold axis in Flock House Virus
(Section~\ref{sec:introduction:symmetry}),
inaccurate results near symmetry axes greatly reduce the value of the
entire experiment.
\par
Beyond correcting the problem in existing computations, this is the first
publication of which the authors are aware in which a full 6-D covariance
function is computed for particles with symmetry.
Interpreting the 6-D information is challenging but a simple example
concerning allosteric interactions is described in
Figure~\ref{fig:SymStat:allosteric}.
We are working with our biological collaborators to extend these
ideas.
\appendices
\section{Constraint on the mean function of $c_{p,\zeta}(x)$}
\label{sec:mean:derivation}
Eq.~(\ref{eq:barc:constraint:xdependent}) is derived in this section.
Define $\bar\rho(\vx)=\expect[\rho(\vx)]\in\mathbb{R}^1$ and
$\bar c_{p,\zeta}(x)=\expect[c_{p,\zeta}(x)]\in\mathbb{R}^{d_p}$.
From (\ref{eq:I2rho}),
\begin{equation}
\bar\rho(\vx)
=
\sum_{p=1}^{\Ngroup}
\sum_\zeta
\bar c_{p,\zeta}^T(x)
I_{p,\zeta}(\vx/x)
\label{eq:barrho}
\end{equation}
and from
(\ref{eq:barrho}) and (\ref{eq:transformationEqn})
($g\in\{1,\dots,\Ngroup\}$)
\begin{equation}
\bar\rho(R_g^{-1} \vx)
=
\sum_{p=1}^{\Ngroup}
\sum_\zeta
\bar c_{p,\zeta}^T(x)
(\Gamma^p(g))^T
I_{p,\zeta}(\vx/x)
\label{eq:rotatedbarrho}
.
\end{equation}
From (\ref{eq:symmetry:firstorder})
it follows that the left hand sides
of
(\ref{eq:barrho}) and (\ref{eq:rotatedbarrho})
are equal so that for
all $g\in\{1,\dots,\Ngroup\}$,
\begin{equation}
\sum_{p=1}^{\Ngroup}
\sum_\zeta
\bar c_{p,\zeta}^T(x)
\left[
I_{d_p}
-
(\Gamma^p(g))^T
\right]
I_{p,\zeta}(\vx/x)
=
0
\label{eq:barc:intermediateresult}
.
\end{equation}
Multiplying by $I_{\pp,\zetap}^T(\vx/x)$ on the right and integrating over the surface of
the sphere implies [via the orthogonality of the $I_{p,\zeta}(\vx/x)$
functions (i.e., (\ref{eq:orthonormalityofangularbasisfunctions}))]
that
for all $g\in\{1,\dots,\Ngroup\}$,
\begin{equation}
\bar c_{\pp,\zetap}^T(x)
\left[
I_{d_{\pp}}
-
(\Gamma^{\pp}(g))^T
\right]
=
0_{1,d_{\pp}}
.
\end{equation}
Taking transposes and renaming $\pp$ to $p$ and $\zetap$ to $\zeta$ gives
that for all $g\in\{1,\dots,\Ngroup\}$,
\begin{equation}
\left[
I_{d_p}
-
\Gamma^p(g)
\right]
\bar c_{p,\zeta}(x)
=
0_{d_p,1}
\label{eq:barc:xdependent}
.
\end{equation}
The identity irrep is the $p=1$ case and for this case, $d_{p=1}=1$ and
$\Gamma^{p=1}(g)=1$ for all $g\in\{1\dots,\Ngroup\}$ so that
$\bar c_{p=1,\zeta}(x)$ is unconstrained by (\ref{eq:barc:xdependent}).
Equation (\ref{eq:barc:xdependent})
implies that for all
$g\in\{1,\dots,\Ngroup\}$,
\begin{equation}
\Gamma^p(g)
\bar c_{p,\zeta}(x)
=
\bar c_{p,\zeta}(x)
\label{eq:barc:xdependent:invariantsubspace}
.
\end{equation}
Consider $p\in\{2,\dots,\Nirrep\}$ such that $d_p>1$.
If $\bar c_{p,\zeta}(x)\neq 0_{d_p,1}$ then there is at least a
one-dimensional subspace that is invariant under the action of
$\Gamma^p(g)$ for all $g\in\{1,\dots,\Ngroup\}$ and therefore, by
definition~\cite[p.~67]{Miller1972}, $\Gamma^p(g)$ is not an irrep.
But this contradicts the assumption that $\Gamma^p(g)$ is an irrep.
Therefore, the statement $\bar c_{p,\zeta}(x)\neq 0_{d_p,1}$ must be false.
Finally, consider $p\in\{2,\dots,\Nirrep\}$ such that $d_p=1$.
If $\bar c_{p,\zeta}(x)\neq 0$ then
(\ref{eq:barc:xdependent:invariantsubspace})
implies that $\Gamma^p(g)=1$
for all $g\in\{1,\dots,\Ngroup\}$ but this is the identity irrep that is
the $p=1$ irrep.
Therefore, $\bar c_{p,\zeta}(x)=0$.
These results are summarized in
(\ref{eq:barc:constraint:xdependent}).
\section{Constraint on the covariance function of $c_{p,\zeta}(x)$}
\label{sec:covariance:derivation}
Eq.~(\ref{eq:Cc:constraint:xdependent}) is derived in this
section acting as if $c_{p,\zeta}(x)$, $I_{p,\zeta}(\vx/x)$, and
$\Gamma^p(g)$ are complex valued, which simplifies the presentation in
Appendix~\ref{sec:constraintsonthemomentsoftheweights:complex}.
Let $\calM_p$ be the range of the ordered pair $\zeta=(l,n)$ in the indices of
$I_{p,j;l,n}(\vx/x)$, which depends on the value of $p$, e.g.,
$\zeta,\zeta_1,\zeta_2\in\calM_p$.
Define
$\tilde\rho(\vx)=\rho(\vx)-\bar\rho(\vx)$
and
$\tilde c_{p,\zeta}(x)=c_{p,\zeta}(x)-\bar c_{p,\zeta}$.
Define
$\CrhoREAL(\vx_1,\vx_2)=\expect[(\rho(\vx_1)-\bar\rho(\vx_1))(\rho(\vx_2)-\bar\rho(\vx_2))]=\expect[\tilde\rho(\vx_1)\tilde\rho(\vx_2)]\in\mathbb{C}^1$
(as in Section~\ref{sec:symmetrysymmetricstatistics})
and
$\CcREAL(x_1,x_2)=\expect[(c_{p_1,\zeta_1}(x_1)-\bar c_{p_1,\zeta_1}(x_1))
(c_{p_2,\zeta_2}(x_2)-\bar c_{p_2,\zeta_2}(x_2))^T]
=\expect[\tilde c_{p_1,\zeta_1}(x_1)\tilde c_{p_2,\zeta_2}^T(x_2)]
\in\mathbb{C}^{d_{p_1}\times d_{p_2}}$.
Equation (\ref{eq:I2rho})
implies that
\begin{eqnarray}
\tilde\rho(\vx)
&=&
\sum_{p=1}^{\Ngroup}
\sum_\zeta
I_{p,\zeta}^T(\vx/x)
\tilde c_{p,\zeta}(x)
\label{eq:I2tilderho:withtranspose}
.
\end{eqnarray}
From (\ref{eq:I2tilderho:withtranspose})
it follows that
\begin{align}
\!\begin{aligned}[t]\CrhoREAL(\vx_1,\vx_2)
=
\sum_{p_1=1}^{\Nirrep}
&\sum_{\zeta_1}
\sum_{p_2=1}^{\Nirrep}
\sum_{\zeta_2}
I_{p_1,\zeta_1}^T(\vx_1/x_1)\\
&{}\times\CcREAL(x_1,x_2) I_{p_2,\zeta_2}(\vx_2/x_2).
\end{aligned}
\label{eq:Crho}
\end{align}
and from
(\ref{eq:Crho}) and (\ref{eq:transformationEqn})
($g\in\{1,\dots,\Ngroup\}$) it follows that
\begin{align}
\!\begin{aligned}[t]\CrhoREAL&(R_g^{-1} \vx_1,R_g^{-1} \vx_2)
={\sum_{p_1=1}^{\Nirrep}
\sum_{\zeta_1}
\sum_{p_2=1}^{\Nirrep}
\sum_{\zeta_2}
I_{p_1,\zeta_1}^T(\vx_1/x_1)}\\
&{}\times\Gamma^{p_1}(g){\CcREAL(x_1,x_2)
[\Gamma^{p_2}(g)]^T
I_{p_2,\zeta_2}(\vx_2/x_2)}.
\end{aligned}
\label{eq:rotatedCrho}
\end{align}
From (\ref{eq:symmetry:secondorder})
it follows that the left hand sides
of (\ref{eq:Crho}) and~(\ref{eq:rotatedCrho})
are equal so that for
all $g\in\{1,\dots,\Ngroup\}$,
\begin{align}
\begin{aligned}[t]
&{\sum_{p_1=1}^{\Nirrep}}
{\sum_{\zeta_1}}
{\sum_{p_2=1}^{\Nirrep}
\sum_{\zeta_2}}
\medmath{I_{p_1,\zeta_1}^T(\vx_1/x_1)}\medmath{\biggl[
\CcREAL(x_1,x_2)}\\
&{~}-\medmath{
\Gamma^{p_1}(g)}
\medmath{\CcREAL(x_1,x_2)
[\Gamma^{p_2}(g)]^T
\biggr]}\medmath{I_{p_2,\zeta_2}(\vx_2/x_2)}
=
0
.
\end{aligned}
\end{align}
Multiplying on the left by $I_{p_1^\prime,\zeta_1^\prime}^\ast(\vx_1/x_1)$ and
on the right by $I_{p_2^\prime,\zeta_2^\prime}^H(\vx_2/x_2)$ and
integrating over the surface of the sphere for both $\vx_1$ and $\vx_2$
implies [via the orthogonality of the $I_{p,\zeta}(\vx/x)$ functions
(i.e., (\ref{eq:orthonormalityofangularbasisfunctions})
with
$I_{\pp,\lp,\np}^T(\vx/x)$ replaced by $I_{\pp,\lp,\np}^H(\vx/x)$)] that for all
$g\in\{1,\dots,\Ngroup\}$,
\begin{align}
\CcprimeREAL(x_1,x_2)=
\Gamma^{p_1^\prime}(g)
\CcprimeREAL(x_1,x_2)
[\Gamma^{p_2^\prime}(g)]^T.
\end{align}
Renaming $p_1^\prime$, $\zeta_1^\prime$, $p_2^\prime$, and $\zeta_2^\prime$
to $p_1$, $\zeta_1$, $p_2$, and $\zeta_2$, respectively, multiplying on the
right by $[\Gamma^{p_2}(g)]^\ast$, taking advantage of the unitarity of
the irrep matrices $\Gamma^p(g)$, and moving the second term to the right
hand side of the equation gives
\begin{equation}
\CcREAL(x_1,x_2)
[\Gamma^{p_2}(g)]^\ast
=
\Gamma^{p_1}(g)
\CcREAL(x_1,x_2)
\label{eq:constraint:covariance:rhorho}
,
\end{equation}
which must be true for all $g\in\{1,\dots,\Ngroup\}$,
$p_1,p_2\in\{1,\dots,\Nirrep\}$, $\zeta_1,\zeta_2\in\calM_p$, and
$x_1,x_2\in\nonnegative$.
Finally, use the fact that the irreps $\Gamma^p(g)$
($p\in\{1,\dots,\Nirrep\}$, $g\in\{1,\dots,\Ngroup\}$) are real-valued
orthonormal rather than complex-valued unitary to get
\begin{equation}
\CcREAL(x_1,x_2)
\Gamma^{p_2}(g)
=
\Gamma^{p_1}(g)
\CcREAL(x_1,x_2)
\label{eq:constraint:covariance:rhorho:Gammareal}
,
\end{equation}
which must be true for all $g\in\{1,\dots,\Ngroup\}$,
$p_1,p_2\in\{1,\dots,\Nirrep\}$, $\zeta_1,\zeta_2\in\calM_p$, and
$x_1,x_2\in\nonnegative$.
This set of matrix equations, each of dimension $d_{p_1}\times
d_{p_2}$, has substantial structure because the $\Gamma^p(g)$
($p_1,p_2\in\{1,\dots,\Nirrep\}$, $g\in\{1,\dots,\Ngroup\}$) matrices are
irreps that are distinct for $p_1\neq p_2$.
Schur's Lemma~\cite[Corollary~3.2, p.~70]{Miller1972}
implies (\ref{eq:Cc:constraint:xdependent}).
\section{$\bar c_{p,\zeta,q}$ and $\mathbf{V}_{p_1,\zeta_1,q_1;p_2,\zeta_2,q_2}$}
\label{sec:finitedimensional}
This section contains detailed notation and derivations for a
finite-dimensional description of the mean and covariance of
$c_{p,\zeta}(x)$ in terms of the mean vector and covariance matrix of the
coefficients in an orthonormal expansion of $c_{p,\zeta}(x)$.
The expansion is
\begin{equation}
c_{p,\zeta}(x)
=
\sum_{q=1}^{N_q}
c_{p,\zeta,q}
\psi_{p,\zeta,q}(x)
\label{eq:radialexpansion}
\end{equation}
where
$c_{p,\zeta,q}\in\mathbb{R}^{d_p}$,
$\psi_{p,\zeta,q}(x)\in\mathbb{R}$
(the radial basis functions of Section~\ref{sec:BasisFunction}),
and
$\int_{x=0}^\infty
\psi_{p,\zeta,q_1}(x)
\psi_{p,\zeta,q_2}(x)
x^2 \dd x
=
\delta_{q_1,q_2}
$.
The goal is to determine the constraints on the mean vector
$\bar c_{p,\zeta,q}$ and covariance matrix $\mathbf{V}_{p_1,\zeta_1,q_1;p_2,\zeta_2,q_2}$
of the weights $c_{p,\zeta,q}$ where
$\bar c_{p,\zeta,q}=\expect[c_{p,\zeta,q}]$
and
$\mathbf{V}_{p_1,\zeta_1,q_1;p_2,\zeta_2,q_2}=\expect[
(c_{p_1,\zeta_1,q_1}-\bar c_{p_1,\zeta_1,q_1})
(c_{p_2,\zeta_2,q_2}-\bar c_{p_2,\zeta_2,q_2})^T]$.
In this notation,
\begin{equation}
\bar c_{p,\zeta}(x)
=
\sum_{q=1}^{N_q}
\bar c_{p,\zeta,q}
\psi_{p,\zeta,q}(x)
\label{eq:meanofradialexpansion}
.
\end{equation}
To determine the mean vector $\bar c_{p,\zeta,q}$,
substitute (\ref{eq:meanofradialexpansion})
into
(\ref{eq:barc:constraint:xdependent}),
multiply by $\psi_{p,\zeta,\qp}(x) x^2$, integrate over
$x\in[0,\infty)$, and rename $\qp$ to $q$ to find
(\ref{eq:barc:constraint:xindependent}).
\par
To determine the covariance matrix $\mathbf{V}_{p_1,\zeta_1,q_1;p_2,\zeta_2,q_2}$,
first note from the definition
$\CcREAL(x_1,x_2)=\expect[(c_{p_1,\zeta_1}(x_1)-\bar c_{p_1,\zeta_1}(x_1))
(c_{p_2,\zeta_2}(x_2)-\bar c_{p_2,\zeta_2}(x_2))^T]$
that (via
(\ref{eq:radialexpansion}) and~(\ref{eq:meanofradialexpansion}))
\begin{align}
\begin{aligned}[t]
\CcREAL(x_1,x_2)=
&\sum_{q_1=1}^{N_q}
\sum_{q_2=1}^{N_q}
\psi_{p_1,\zeta_1,q_1}(x_1)\\
&{}\times \mathbf{V}_{p_1,\zeta_1,q_1;p_2,\zeta_2,q_2}
\psi_{p_2,\zeta_2,q_2}(x_2),
\end{aligned}
\label{eq:covarianceofradialexpansion}
\end{align}
so that
\begin{align}
\begin{aligned}[t]
&\mathbf{V}_{p_1,\zeta_1,q_1^\prime;p_2,\zeta_2,q_2^\prime}
=
\int_{x_1=0}^\infty
\int_{x_2=0}^\infty
\psi_{p_1,\zeta_1,q_1^\prime}(x_1)\\
&{~~~~~~}\times\CcREAL(x_1,x_2)
\psi_{p_2,\zeta_2,q_2^\prime}(x_2)
x_1^2
x_2^2
\dd x_1
\dd x_2.
\end{aligned}
\label{eq:VfromCc}
\end{align}
Define $v_{p_1}(\zeta_1,q_1;\zeta_2,q_2)\in\mathbb{R}$ by
\begin{align}
\begin{aligned}[t]
v_{p_1}(&\zeta_1,q_1;\zeta_2,q_2)
=
\int_{x_1=0}^\infty
\int_{x_2=0}^\infty
\psi_{p_1,\zeta_1,q_1}(x_1)\\
&{~~~~}\times
c_{p_1}(\zeta_1,x_1;\zeta_2,x_2)
\psi_{p_1,\zeta_2,q_2}(x_2)
x_1^2
x_2^2
\dd x_1
\dd x_2.
\end{aligned}
\end{align}
Substitute (\ref{eq:Cc:constraint:xdependent})
into the right hand side
of (\ref{eq:VfromCc})
and then replace $q_1^\prime$ and $q_2^\prime$ by
$q_1$ and $q_2$, respectively, to find (\ref{eq:Cc:constraint:xindependent}).
\vspace*{-.07in}
\section{Cases where real-valued irreps are not available}
\label{sec:constraintsonthemomentsoftheweights:complex}
Standard references for rotational point group irreps give the irrep
matrices in complex-valued unitary form.
Frobenious-Schur theory~\cite[p.~129, Theorem~III]{Cornwell1984} implies
that
among $I$, $O$, $T$, and $C_n$
only for $I$, $O$, $C_1$, and $C_2$ does there exist a similarity
transformation to a real-valued orthonormal form for all irreps of the group.
In this section, the constraints on the moments of the weights caused by
symmetric statistics are investigated when the irreps and therefore the
basis functions and the weights are complex.
\par
\par
The results for the mean function of $c_{p,\zeta}(x)$ are unaltered from
(\ref{eq:barrho})--(\ref{eq:barc:xdependent:invariantsubspace}) and
(\ref{eq:barc:constraint:xdependent}).
\par
The results for the covariance function of $c_{p,\zeta}(x)$ are changed
from
(\ref{eq:I2tilderho:withtranspose})--(\ref{eq:constraint:covariance:rhorho}).
Because the variables are now complex, it is necessary to compute and to
apply symmetry constraints to both
$\Crhorho(\vx_1,\vx_2)$ and $\Crhorhoconj(\vx_1,\vx_2)$, which
are defined by
$\Crhorho(\vx_1,\vx_2)=\expect[(\rho(\vx_1)-\bar\rho(\vx_1))(\rho(\vx_2)-\bar\rho(\vx_2))]$
(the same as $\CrhoREAL(\vx_1,\vx_2)$ in Section~\ref{sec:constraintsonthemomentsoftheweights})
and
$\Crhorhoconj(\vx_1,\vx_2)=\expect[(\rho(\vx_1)-\bar\rho(\vx_1))(\rho(\vx_2)-\bar\rho(\vx_2))^\ast]$,
respectively.
The calculations for $\Crhorho$ are the same as
(\ref{eq:I2tilderho:withtranspose})--(\ref{eq:constraint:covariance:rhorho}).
In a more detailed notation where
$\bar c_{p,\zeta}(x)=\expect[c_{p,\zeta}(x)]$
and
$\Ccc(x_1,x_2)=\expect[(c_{p_1,\zeta_1}(x_1)-\bar c_{p_1,\zeta_1}(x_1))
(c_{p_2,\zeta_2}(x_2)-\bar c_{p_2,\zeta_2}(x_2))^T]$,
the result is
\begin{equation}
\Ccc(x_1,x_2)
[\Gamma^{p_2}(g)]^\ast
=
\Gamma^{p_1}(g)
\Ccc(x_1,x_2)
\label{eq:constraint:covariance:rhorho:2}
.
\end{equation}
It is not possible to continue on to
(\ref{eq:constraint:covariance:rhorho:Gammareal})
because $\Gamma^p(g)$
is no longer real and a new solution is described in the following
paragraph.
The calculations for $\Crhorhoconj$ follow the same plan with the
result that
[$\Cccconj(x_1,x_2)=\expect[(c_{p_1,\zeta_1}(x_1)-\bar c_{p_1,\zeta_1}(x_1))
(c_{p_2,\zeta_2}^\ast(x_2)-\bar c_{p_2,\zeta_2}^\ast(x_2))^T]$]
\begin{equation}
\Cccconj(x_1,x_2)
\Gamma^{p_2}(g)
=
\Gamma^{p_1}(g)
\Cccconj(x_1,x_2)
\label{eq:constraint:covariance:rhorhoconj}
.
\end{equation}
Equation (\ref{eq:constraint:covariance:rhorhoconj})
is of the same form as
(\ref{eq:constraint:covariance:rhorho:Gammareal})
and so has the same form of solution, which is
\begin{align}
\Cccconj&(x_1,x_2)\nonumber\\
=&
\left\{
\begin{array}{ll}
c_{p_1}^{c,c^\ast}(\zeta_1,x_1;\zeta_2,x_2)
I_{d_{p_1}}
,
&
p_1=p_2
\\
0_{d_{p_1},d_{p_2}}
,
&
\mbox{otherwise}
\end{array}
\right.
\label{eq:Cccconj:constraint:xdependent}
.
\end{align}
\par
The solution to (\ref{eq:constraint:covariance:rhorho:2})
is described in
this paragraph.
Two irreps are equivalent~\cite[Thm.~I, p.~71]{Cornwell1984}
when there exists an invertible matrix $S$ such
that $\Gamma_1(g)=S^{-1}\Gamma_2(g)S$ for all $g$.
Since $\Gamma^{p_2}(g)$ is an irrep, it follows that
$[\Gamma^{p_2}(g)]^\ast$ is a rep.
By Ref.~\cite[Theorem~3.5 p.~70]{Miller1972}, $[\Gamma^{p_2}(g)]^\ast$ is
an irrep.
There is a fixed set of irreps, so there must be a function $\tau$ from
$\{1,\dots,\Nirrep\}$ to $\{1,\dots,\Nirrep\}$ such that
$[\Gamma^{p_2}(g)]^\ast=S^{-1} \Gamma^{\tau(p_2)}(g) S$ for some invertible
matrix $S$.
The function $\tau$ must be a permutation since if
$\tau(p_2)=\tau(p_2^\prime)=p_\ast$ then
$S [\Gamma^{p_2}(g)]^\ast S^{-1}=\Gamma^{p_\ast}(g)$
and
$S^\prime [\Gamma^{p_2^\prime}(g)]^\ast [S^\prime]^{-1}=\Gamma^{p_\ast}(g)$.
Equating the left hand sides of these two equations gives
$S [\Gamma^{p_2}(g)]^\ast S^{-1}=S^\prime [\Gamma^{p_2^\prime}(g)]^\ast [S^\prime]^{-1}$
which implies
$T^{-1} \Gamma^{p_2}(g) T = \Gamma^{p_2^\prime}(g)$
(where $T=[S^{-1} S^\prime]^\ast$)
which implies that $p_2$ and $p_2^\prime$ are the same irrep.
Therefore, (\ref{eq:constraint:covariance:rhorho:2})
is equivalent to
\begin{equation}
\Ccc(x_1,x_2)
\Gamma^{\tau(p_2)}(g)
=
\Gamma^{p_1}(g)
\Ccc(x_1,x_2)
\label{eq:constraint:covariance:rhorho:3}
.
\end{equation}
Equation (\ref{eq:constraint:covariance:rhorho:3})
is of the same form as
(\ref{eq:constraint:covariance:rhorho:Gammareal})
and so has the same
form of solution, which is
\begin{align}
\Ccc&(x_1,x_2)\nonumber\\
=&
\left\{
\begin{array}{ll}
c_{p_1}^{c,c}(\zeta_1,x_1;\zeta_2,x_2)
I_{d_{p_1}}
,
&
p_1=\tau(p_2)
\\
0_{d_{p_1},d_{\tau(p_2)}}
,
&
\mbox{otherwise}
\end{array}
\right.
\label{eq:Ccc:constraint:xdependent}
.
\end{align}
\par
Rather than $\Ccc(x_1,x_2)$ and
$\Cccconj(x_1,x_2)$, it is probably more
natural to work with real-valued functions.
This can be achieved by working with the real and imaginary parts of
$c_{p,\zeta}(x)-\bar c_{p,\zeta}(x)$ leading by standard calculations to
formulas for
$\CRecRec(x_1,x_2)$, $\CImcImc(x_1,x_2)$, and $\CRecImc(x_1,x_2)$.
Either
$\Ccc(x_1,x_2)$ and
$\Cccconj(x_1,x_2)$
or
$\CRecRec(x_1,x_2)$,
$\CImcImc(x_1,x_2)$, and
$\CRecImc(x_1,x_2)$
can be made finite-dimensional by the same methods used in
Section~\ref{sec:constraintsonthemomentsoftheweights}.
\section{Use of non-spherical coordinate systems}
\label{sec:nonsphericalcoordinatesystems}
Except in this section, results in this paper are stated in the spherical
coordinate system (e.g., (\ref{eq:I2rho})).
However, for particles obeying the symmetries of $C_n$ (and in some cases
$D_n$)
cylindrical coordinates
$(r,\phi,z)$ can be more natural with the single symmetry axis of $C_n$ on
the $z$ axis of the coordinate system.
For the case of $C_n$, (\ref{eq:I2rho})
is replaced by
$\rho(\vx)
=
\sum_{p=1}^{\Ngroup}
\sum_\zeta
c_{p,\zeta}^T(r,z)
I_{p,\zeta}(\phi)$
and the constraint of symmetric statistics specifies the mean and
covariance functions of $c_{p,\zeta}(r,z)$ rather than the mean
(see (\ref{eq:barc:constraint:xdependent}))
and covariance
(see (\ref{eq:Cc:constraint:xdependent}) or (\ref{eq:Cccconj:constraint:xdependent}) and~(\ref{eq:Ccc:constraint:xdependent}))
functions of $c_{p,\zeta}(x)$.
In either case, the effect of the symmetric statistics constraint is a
constraint on the covariance function of the coordinate-dependent weights
where the weights depend on the coordinates that are not involved in the
symmetry operation.
\section*{Acknowledgments}
We are grateful to Profs.\ John E. Johnson (The Scripps Research
Institute) and David Veesler (Univ.\ of Washington) for the HK97 data
and helpful discussions;
to Prof.\ Johnson and Dr.\ Tsutomu Matsui (Stanford University) for the
N$\omega$V data and helpful discussions;
to Drs.\ Yili Zheng, Qiu Wang, and Yunye Gong for
their effort on the {\tt Hetero} software;
and to NSF 1217867 for support.
\ifCLASSOPTIONcaptionsoff
  \newpage
\fi
\bibliographystyle{IEEEtran}
\bibliography{../../../references}
\vspace*{-1cm}
\begin{IEEEbiography}[{\includegraphics[width=1in,height=1.25in,clip,keepaspectratio]{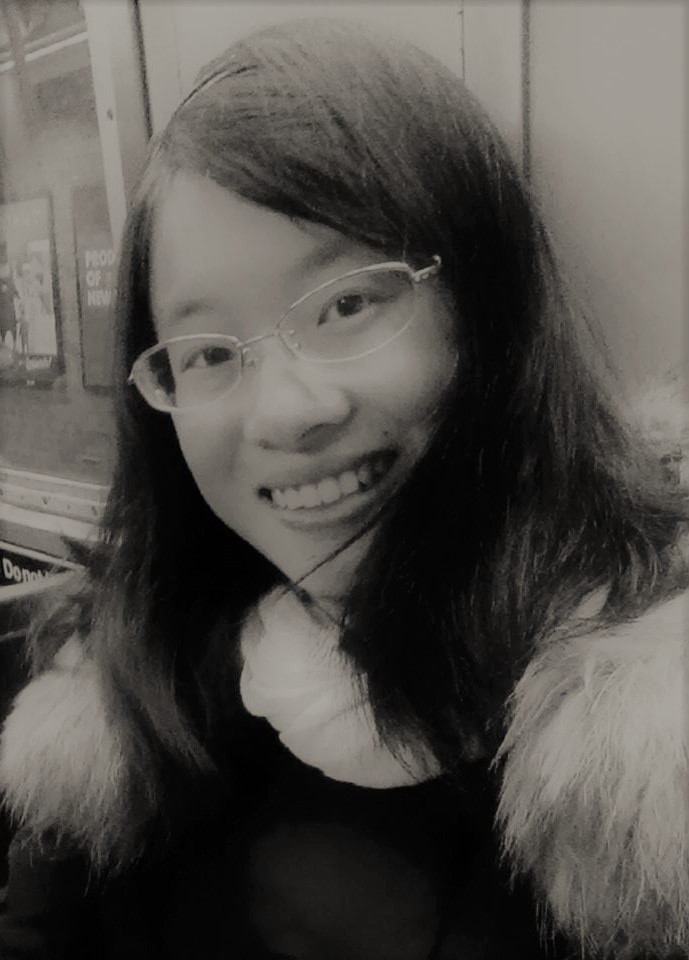}}]%
{Nan Xu} received her Ph.D. degree from the School of Electrical and
Computer Engineering with a minor in Applied Mathematics and a minor in
Cognitive Neuroscience at Cornell University in May 2017. Nan received her Master of Science degree in May 2015. Before joining Cornell, she received double Bachelor's degrees in Electrical and Computer Engineering (BS) and Mathematics (BA) with a minor in Music at the University of Rochester in Rochester NY. Nan's current research interests are in statistical modeling and inference in biological data. She is investigating three primary applications: brain network estimation, realistic fMRI data simulation, and reconstruction of the statistical characteristics of ensembles of heterogeneous virus particles.
\end{IEEEbiography}
\begin{IEEEbiography}[{\includegraphics[width=1in,height=1.25in,clip,keepaspectratio]{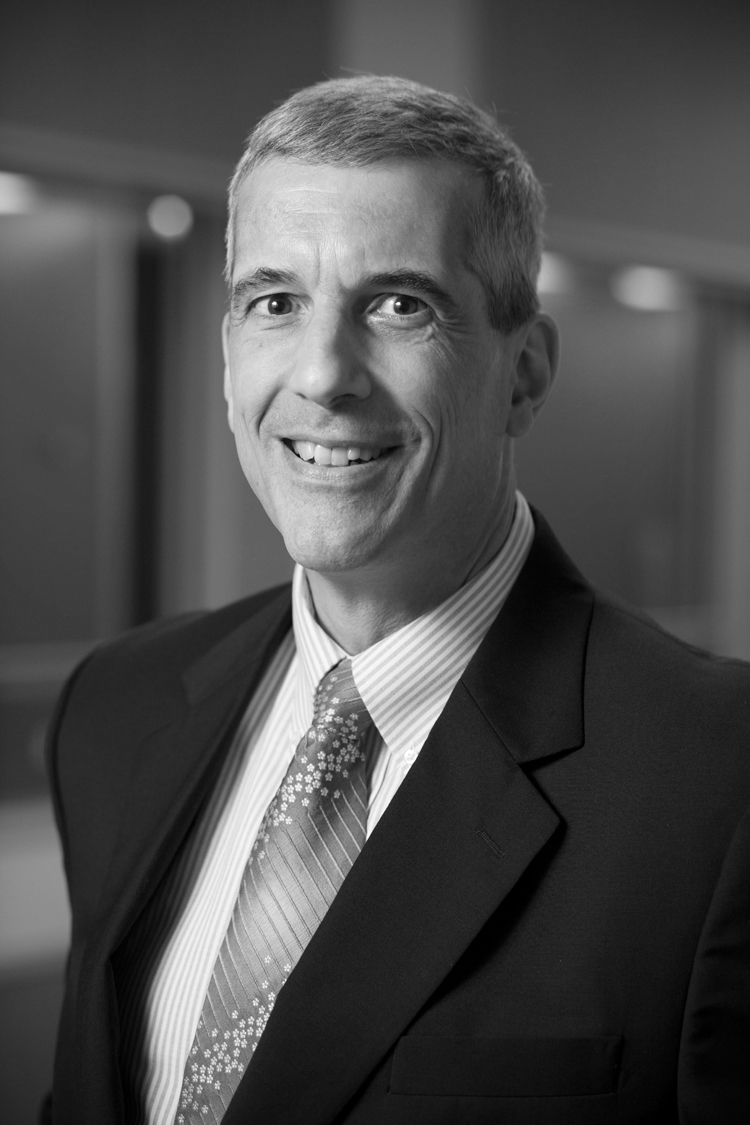}}]%
{Peter C. Doerschuk}
(SM'03) received the B.S.,
M.S., and Ph.D. degrees in electrical engineering
from the Massachusetts Institute of Technology
(MIT), Cambridge, in 1977, 1979, and 1985, respectively,
and the M.D. degree from Harvard Medical
School, Cambridge, MA, in 1987.
After postgraduate training at Brigham and
Womens' Hospital, he held a postdoctoral appointment
with the Laboratory for Information
and Decision Systems, MIT, from January 1988
to August 1990. After 16 years on the faculty in
Electrical and Computer Engineering, Purdue University, West Lafayette, IN,
he joined the faculty in Biomedical Engineering and Electrical and Computer
Engineering, Cornell University, Ithaca, NY, in July 2006.
\end{IEEEbiography}
\end{document}